\documentclass[fleqn,usenatbib]{mnras}

\usepackage[T1]{fontenc}

\DeclareRobustCommand{\VAN}[3]{#2}
\let\VANthebibliography\thebibliography
\def\thebibliography{\DeclareRobustCommand{\VAN}[3]{##3}\VANthebibliography}

\usepackage{graphicx}	
\usepackage{amsmath}	
\usepackage{amssymb}	
\usepackage{caption}
\usepackage{subcaption}
\usepackage{dcolumn}
\usepackage{bm}
\usepackage{hyperref}
\usepackage{cleveref}
\usepackage[dvipsnames]{xcolor}
\usepackage{soul}
\Crefformat{figure}{#2Fig.~#1#3}
\Crefmultiformat{figure}{Figs.~#2#1#3}{ and~#2#1#3}{, #2#1#3}{ and~#2#1#3}
\Crefformat{equation}{#2Eq.~(#1)#3}
\Crefmultiformat{equation}{Eqs.~#2(#1)#3}{ and~#2(#1)#3}{, #2(#1)#3}{ and~#2(#1)#3}

\definecolor{ForestGreen}{rgb}{0.2,0.6,0.2}

\newcommand{\muscle}{{\scshape muscle}}
\newcommand{\muscleups}{{\scshape muscle-ups}}
\newcommand{\origami}{{\scshape origami}}
\newcommand{\rockstar}{{\scshape rockstar}}
\newcommand{\peakpatch}{{\scshape peak-patch}}
\newcommand{\hmpc}{{$h^{-1}$Mpc}}
\newcommand{\mpch}{{$h$ Mpc$^{-1}$}}

\title[MUSCLE-UPS: EPS+LPT Improved Simulations]{MUSCLE-UPS: Improved Approximations of the Matter Field with the Extended Press-Schechter Formalism and Lagrangian Perturbation Theory}

\author[F. Tosone et al.]{Federico Tosone,$^{1}$\thanks{E-mail: federico.tosone@unimi.it}
Mark C.\ Neyrinck,$^{4,5,6}$
Benjamin R. Granett,$^{1,7,8}$
\newauthor
Luigi Guzzo,$^{1,7,8}$
Nicola Vittorio$^{2,3}$
\\
$^{1}$Dipartimento di Fisica, Universit\`{a} degli Studi di Milano, via G. Celoria 16, 20133 Milano, Italy\\
$^{2}$Dipartimento di Fisica, Universit\`{a} di Roma Tor Vergata, Via della Ricerca Scientifica 1, I-00133, Roma, Italy\\
$^{3}$INFN Sezione di Roma Tor Vergata, Via della Ricerca Scientifica 1, I-00133, Roma, Italy\\
$^{4}$Ikerbasque, the Basque Foundation for Science\\
$^{5}$Dept.\ of Physics, University of the Basque Country UPV/EHU, Bilbao, Spain\\
$^{6}$Donostia International Physics Center, San Sebasti\'{a}n, Spain\\
$^{7}$INAF - Osservatorio Astronomico di Brera, via Brera 28, 20122 Milano, Italy\\
$^{8}$INFN Sezione di Milano, Via G. Celoria 16, 20133, Milano, Italy
}

\date{Accepted XXX. Received YYY; in original form ZZZ}

\pubyear{2021}

\begin{document}
\label{firstpage}
\pagerange{\pageref{firstpage}--\pageref{lastpage}}
\maketitle

\setstcolor{red}
\begin{abstract}
Lagrangian algorithms to simulate the evolution of cold dark matter (CDM) are invaluable tools to generate large suites of mock halo catalogues. In this paper, we first show that the main limitation of current semi-analytical schemes to simulate the displacement of CDM is their inability to model the evolution of overdensities in the initial density field, a limit that can be circumvented by detecting halo particles in the initial conditions. We thus propose `MUltiscale Spherical Collapse Lagrangian Evolution Using Press-Schechter' (\muscleups), a new scheme that reproduces the results from Lagrangian perturbation theory on large scales, while improving the modelling of overdensities on small scales. In \muscleups, we adapt the extended Press and Schechter (EPS) formalism to Lagrangian algorithms of the displacement field. For regions exceeding a collapse threshold in the density smoothed at a radius $R$, we consider all particles within a radius $R$ collapsed. Exploiting a multi-scale smoothing of the initial density, we build a halo catalogue on the fly by optimizing the selection of halo candidates. This allows us to generate a density field with a halo mass function that matches one measured in $N$-body simulations. We further explicitly gather particles in each halo together in a profile, providing a numerical, Lagrangian-based implementation of the halo model. Compared to previous semi-analytical Lagrangian methods, we find that \muscleups\ improves the recovery of the statistics of the density field at the level of the probability density function (PDF), the power spectrum, and the cross correlation with the $N$-body result.
\end{abstract}


\section{Introduction}
Structures in the Universe formed through gravitational clustering around perturbations in an almost homogeneous initial matter distribution. These seeds eventually became haloes, gravitationally bound structures which set the stage for the formation of the cosmic web wrapped around them. 

An insight into this process was offered by the Lagrangian picture, where one traces the evolution of each particle from its initial position to its final, Eulerian position. If CDM is treated as an ideal, pressureless and self-gravitating fluid, the formation of caustics is expected \citep{Shandarin89,Buchert92}. This was also confirmed by numerical simulations based on the same formalism \citep{Bouchet1995}: perturbative schemes such as the Zel'dovich approximation (ZA) and second-order Lagrangian perturbation theory (2LPT) are so effective in reproducing the cosmic web \citep{Bond95}, that they became customary techniques for the generation of mock galaxy catalogues. They are adopted by many codes such as \peakpatch\ \citep{Bond96,Stein18}, \textsc{pinocchio} \citep{Monaco01,Monaco13}, \textsc{PThaloes} \citep{Scoccimarro02}, \textsc{cola} \citep{Tassev13}, \textsc{halogen} \citep{Avila14}, \textsc{patchy} \citep{Kitaura14}, \textsc{EZmocks} \citep{Chuang15a}. The reason behind the success of semi-analytical schemes is that they allowed to build hundreds of simulated realizations for a given cosmology, which is still prohibitive for $N$-body codes. Moreover, these approximate schemes have opened up the path to the reconstruction of initial conditions by forward Bayesian methodologies \citep{Jasche13,Kitaura13,Leclercq15,Bos18} or reconstructions based on reversing gravity evolution \citep{Padmanabhan12}.

Despite their importance and widespread use, the progress in Lagrangian semi-analytical techniques to simulate the displacement field has been slow, due to the intrinsic limitations of perturbation theory. First, it is accurate only for small perturbations, thus failing at low redshift and small scales. Secondly, it moves particles only according to the initial potential, so it becomes inaccurate when caustics occur and particle trajectories overlap. An advance came when \cite{Neyrinck13} proposed a non-perturbative technique based on the spherical collapse model (SC), which provides a better description of voids and overdensities in the simulations. The SC approach can be combined with 2LPT, valid on large scales; the resulting technique was dubbed Augmented Lagrangian Perturbation Theory (ALPT). It performs remarkably well at the level of fundamental statistics of the CDM density field \citep{Kitaura13_ALPT}, and has been extensively used for the generation of mock catalogues, e.g.\ providing the basis for the the widely used \textsc{patchy} mocks for the Baryon Oscillation Spectroscopic Survey \citep{KitauraEtal2016}. Later, \cite{Neyrinck15} proposed a new scheme named \muscle, an improved non-perturbative method to detect halo particles through a multi-scale smoothing process based on the SC criterion. This approach has the benefit of conceptual clarity, and recovers the linear power spectrum much better than SC on large scales, while being comparable to ALPT with respect to the small scale clustering.

The possibility to improve Lagrangian non-perturbative techniques motivated our recent study \citep{Tosone20}, where we measured the statistics of the density field generated from a displacement field that reproduces the same PDF and power spectrum of the exact displacement measured from the $N$-body. We found only marginal benefits, with results not better than \muscle. We concluded that Lagrangian methods based on a mapping of the raw density field exhausted their utility to make further progress, and any improvement should come from a multi-scale approach. This is in fact the subject of the work presented here, which we regard as a step forward, built up on top of existing schemes in the literature.

After briefly recalling the basic approximation schemes in \Cref{sect:theory} and outlining the methodology in \Cref{sect:methods}, in \Cref{sect:limits} we show the reasoning that led to our proposed approach. We start by considering truncated schemes of Lagrangian approximations, as we confirm that they perform as well as non-truncated schemes \citep{Coles93} in terms of power spectrum and cross correlation statistics, even though the overdensities, and consequently the PDF of the density field, are completely off with respect to the N-body \citep[see also][for a discussion on the limits of truncated schemes]{Munari17}. This confirms that the main limitation of semi-analytical techniques is their lack of predictivity for overdense regions, due to shell crossing. We further state our point about the importance of overdensities in \Cref{sect:halofinders}, where we show that a better identification of halo particles in the initial conditions is enough to improve substantially the approximation of the displacement field.

To this end in \Cref{sect:muscleups} we propose a predictive scheme which is based on the EPS formalism \citep{Zentner06} adapted to Lagrangian semi-analytical simulations. A voxel (i.e. a cubic cell on the mesh grid) that exceeds a collapse threshold in the density smoothed at some scale $R$ should collapse along with other voxels within (some multiple of) $R$. We dub this technique MUltiscale Spherical Collapse Lagrangian Evolution Using Press-Schechter (\muscleups). To correctly recover the power spectrum expected from linear theory on large scales, we interpolate the displacement field generated through \muscleups\ on small scales with truncated 2LPT (T2LPT) on large scales, using the same technique of ALPT.

Since \muscleups\ is based on a multi-scale smoothing of the initial conditions, in \S \ref{sect:toyhalo} we propose a method to build a halo catalogue based on the smoothing scale of collapse of each voxel and the density of each proto-halo patch. By exploiting the freedom in the choice of merging events, we can optimize the construction of a halo catalogue by matching it to a target HMF as measured in $N$-body simulations. Finally, we show how to use this halo catalogue to implement the halo model \citep{Peacock00,Seljak00}.

The main result of this new scheme is the ability to improve the recovery of the power spectrum at quasi-linear scales, the recovery of the high-density tail of the PDF, and the cross-correlation with the density field in the quasi-nonlinear regime. We show the results in \Cref{sect:results}, and conclude in \Cref{sect:conclusions}.

\section{Theory}\label{sect:theory}
\subsection{Zel'dovich Approximation}\label{subsect:ZA}
The Lagrangian description of structure formation consists of finding the field $\pmb{\Psi}$ that maps an initial grid of particles, with uniform density $\rho_0$, from their coordinates $\pmb{q}$ to the final Eulerian positions 
\begin{equation}\label{eq:master}
    \pmb{x(q,\tau)} = \pmb{q} + \pmb{\Psi}(\pmb{q},\tau).
\end{equation}

The transformation from Eulerian density to Lagrangian density in the single-stream regime, before shell crossing occurs, must satisfy the condition
\begin{equation}
    \rho(\pmb{x},\tau) d^3 x = \rho_0 d^3 q,
\end{equation}
which can be recast in terms of the Jacobian of the transformation
\begin{equation}
\left\| \frac{d^3 x}{d^3 q} \right\| = J = \frac{1}{1+\delta},
\end{equation}
that is equivalent to
\begin{equation}\label{eq:masselement}
    1+\delta(\pmb{q},\tau)=\frac{1}{\left\| \delta^{(k)}_{ij} +\Psi_{i,j}(\pmb{q},\tau) \right\|},
\end{equation}
where $\Psi_{i,j}$ stands for the $j$-th derivative of the $i$-th component of the displacement vector, and $\delta^{(k)}_{ij}$ is a Kronecker delta.

From \Cref{eq:masselement} one can immediately see that a perturbative treatment involves only the scalar component of $\psi$ at first order, defined as $\psi = \nabla \cdot \pmb{\Psi}$, and it reads
\begin{equation}\label{eq:zeld}
    \psi^{(1)}(\pmb{q},\tau)  = -\delta^{(1)}(\pmb{q},\tau),
\end{equation}
commonly known as the Zel'dovich Approximation (ZA). Assuming that the displacement field is irrotational, $\pmb{\Psi}$ can be expressed as a function of a displacement potential
\begin{equation}\label{eq:irr}
    \pmb{\Psi}(\pmb{q},\tau) = - \nabla_q \phi(\pmb{q},\tau).
\end{equation}
In the ZA we can define a displacement potential $\phi^{(1)}$ that allows us to define a matrix $\Psi^{(1)}_{i,j}=-\phi^{(1)}_{,ij}$, which upon diagonalization yields for \Cref{eq:masselement} (see \Cref{appendix:tidal}) 
\begin{equation}\label{eq:eigs}
    1+\delta^{(1)}(\pmb{q},\tau) = \frac{1}{\left\|(1-\lambda_1)(1-\lambda_2)(1-\lambda_3)\right\|},
\end{equation}
where $\lambda_i$ are the eigenvalues of $\phi^{(1)}_{,ij}$. From this equation it is also clear the limit of perturbation theory also becomes clear: when $\lambda_i \sim 1$, the perturbative regime breaks down.

\subsection{Second-Order Lagrangian Perturbation Theory}
In order to derive the evolution of CDM particles, one must solve the equation of motion. If we approximate CDM particles as a pressure-less self-gravitating system in an expanding Universe, the equation of motion is
\begin{equation}\label{eq:fundamental}
    \frac{d^2}{d \tau^2} \pmb{x} + \mathcal{H} \frac{d}{d \tau} \pmb{x} = -\nabla_{x} \Phi(\pmb{x}),
\end{equation}
where we used the conformal time $d \tau = dt / a$, $\mathcal{H}= H/a$ and $\Phi$ is the peculiar gravitational potential. The subscript $\pmb{x}$ of the nabla operator indicates a gradient in Eulerian space.

One can formally prove that at first order in Lagrangian perturbation theory, the solution consists of a scalar component only, and it is in fact the ZA (\Cref{eq:zeld}). Moreover, the spatial component is separable from the time component, and we can write 
\begin{equation}
    \psi^{(1)}(\pmb{q}, \tau) = - \delta_{\ell}(\pmb{q}) D_1(\tau),
\end{equation}
where $\delta_{\ell}(\pmb{q})= \delta^{(1)}(\pmb{q},\tau)/D_1(\tau) $. The time dependence can be found solving the equation
\begin{equation}\label{eq:D1}
    \frac{d^2}{d \tau^2} D_1(\tau) + \mathcal{H} \frac{d}{d \tau} D_1(\tau) = \frac{3}{2}\Omega_m(\tau)  \mathcal{H}^2 D_1(\tau).
\end{equation}
The solution $D_1$ is the linear growth function, and it is often described by fitting formulae, rather than by an exact solution. A widespread fitting formula for the growth function in a $\Lambda$CDM Universe is \citep{Carrol1992}
\begin{equation}
    D_{1}(a) \simeq \frac{5}{2} \frac{a \Omega_m(a)}{\Omega_m(a)^{4/7}-\Omega_{\Lambda}(a)+(1+\Omega_m(a)/2)(1+\Omega_{\Lambda}(a)/70)}.
\end{equation}

Using a perturbative approach, it is possible to derive the solution at second order in perturbation theory. Just as in the first order, also at second order the spatial and the temporal parts are separable, which allows to find the second order solution
\begin{equation}
    \psi^{(2)}(\pmb{q},\tau) = \frac{D_2(\tau)}{2 D^2_1(\tau)} \sum_{i \neq j} \left( \phi^{(1)}_{i,i} \phi^{(1)}_{j,j} - \phi^{(1)}_{i,j} \phi^{(1)}_{j,i}\right),
\end{equation}
where $\phi^{(1)}$ is the first-order displacement potential, and $D_2$ is the second-order growth function, solution of the temporal part, which is often approximated by the convenient fit \citep{Bouchet1995}
\begin{equation}
    D_2(\tau) \simeq -\frac{3}{7} D^2_1(\tau) \Omega^{-\frac{1}{143}}(\tau)
\end{equation}
To summarize, the displacement in second-order Lagrangian perturbation theory (2LPT) is:
\begin{equation}
    \pmb{\Psi}_{\text{2lpt}}(\pmb{q},\tau) = - \nabla_{q} \phi^{(1)}(\pmb{q},\tau) + \nabla_{q} \phi^{(2)}(\pmb{q},\tau),
\end{equation}
where we defined the second-order displacement potential 
\begin{equation}
    \pmb{\Psi}^{(2)}(\pmb{q},\tau) = \nabla_{q} \phi^{(2)}(\pmb{q},\tau).
\end{equation}
One can push the perturbative treatment to higher orders, but the complexity increases as the scalar component receives a contribution from the curl component \citep{Bouchet1995}.

\subsection{Spherical collapse and MUSCLE}\label{subsect:SC}
A non-perturbative approach to model the scalar displacement field was proposed by \cite{Neyrinck13}, who used a formula derivable from results in \cite{Bernardeau1994b} \citep[see also][]{Mohayaee1996} to approximate the density evolution of isolated volume elements well. The spherical-collapse (SC) approach obtains the displacement divergence
\begin{equation}\label{eq:sc}
\psi_{\text{sc}}(\pmb{q}, \tau)=
\begin{cases}
3\left[ \left(1- D_1(\tau) \frac{\delta_\ell(\pmb{q})}{\gamma} \right)^{\gamma/3} -1. \right],  &\delta_\ell<\gamma/D_1 \\
-3,  &\delta_\ell \geq \gamma/D_1,
\end{cases}
\end{equation}
where $\gamma$ is a parameter which in the limit of $\Omega_{\text{cdm}} \sim 0 $ is exactly $3/2$ \citep{Bernardeau94}. It was also shown that this choice of $\gamma$ extends well to other cosmologies, especially for underdensities.

This $\gamma$ can be thought of as the critical overdensity of collapse $\delta_c$, predicted from linear perturbation theory to be $1.686$ for an Einstein-de Sitter cosmology. For continuity with the original choice of $\gamma$ in \Cref{eq:sc}, and with the works in the literature based on it, we keep it to $1.5$. Coincidentally, in \cite{Stein18} a value of a linear critical overdensity $\delta_c=1.5$ is also used in order to not underestimate the number of candidate haloes. This value seems to work well also in our case, when we need to find halo regions as implemented in our algorithm in \Cref{sect:muscleups}.

Using \Cref{eq:sc} alone results in a power spectrum of the generated density field that is typically offset from the linear power spectrum on linear scales. The reason for this linear-scale offset was made clear in the \muscle\ algorithm; SC misses the void-in-cloud process, which happens when a voxel density is under a collapse threshold on the grid scale, but over the threshold when smoothed on a larger scale. \muscle\ checks for collapse on increasingly larger scales of the linear density field smoothed through a window $W(k)$
\begin{equation} \label{eq:truncate}
	\delta_\ell(\pmb{q},R) = \int \frac{d^3 k}{(2 \pi)^3} e^{-i \pmb{k} \cdot \pmb{q}} \delta_\ell(\pmb{k}) W(\pmb{k},R).
\end{equation}
If $\delta_{\ell}(\pmb{q},R) \geq \gamma/D_1$ at any scale $R$, then we set $\psi(\pmb{q})=-3$. The value of $-3$ is indeed measured in halo particles from $N$-body simulations, and the multi-scale smoothing allows to check for overdense voxels that are expected to collapse at various scales.

\subsection{Augmented Lagrangian Perturbation Theory}
Before the appearance of \muscle, the so called Augmented LPT (ALPT) was proposed to remedy the large-scale power offset in SC more explicitly. This is achieved convolving the 2LPT scalar divergence field through a low-pass filter and the SC formula of \Cref{eq:sc} through a high-pass filter \citep{Kitaura13_ALPT}
\begin{equation}\label{eq:alpt_ansatz}
    \psi_{\text{alpt}}(\pmb{q},\tau) = \psi_{\text{2lpt}}(\pmb{q},\tau)\circledast \mathcal{G(\sigma_R)}+ \psi_{\text{sc}}(\pmb{q},\tau) \circledast (1- \mathcal{G(\sigma_R)}).
\end{equation}

This is an interpolation between 2LPT on linear scales and SC on nonlinear scales. The two regimes are connected through a smoothing Gaussian kernel $\mathcal{G}$, where $\sigma_R$ is a free parameter of the approach, estimated to be around $3$ \hmpc\ by a fit to $N$-body simulations.

In principle, an interpolation between large and small scales can be adopted also with \muscle, but this provides marginal improvement with respect to ALPT at the cost of a slightly greater computational cost due to the multi-scale smoothing required \citep{Munari17}.
For a detailed comparison between fast simulation schemes via the Lagrangian picture, we refer the reader to the reviews \cite{Monaco16} and \cite{Munari17}.

\section{Method}\label{sect:methods}
In this work, we shall compare approximate evolution of particles to the corresponding exact $N$-body results. Our reference $N$-body simulation was run with \textsc{gadget} \citep{Gadget}, starting at an initial redshift of $z=50$ and fiducial cosmology $\Omega_\text{b} h^2 = 0.0225$, $\Omega_{\text{cdm}}=0.25$ and $\sigma_{8}=0.8$. This simulation has box-size $256$ \hmpc\ and a total of $256^3$ particles. The initial conditions of the simulation have been generated through 2LPT, with amplitudes of the power spectrum fixed to the linear theory prediction \citep{Angulo16}. Throughout this work, we use the same initial conditions for the approximate schemes as well, so that we can quantify their agreement with the reference $N$-body result at the level of a single realization.

Given any prescription for the scalar displacement field $\psi$, we can always connect it to the displacement potential by taking the divergence of \Cref{eq:irr}
\begin{equation}
    \psi(\pmb{q},\tau) =  - \nabla^2_q \phi(\pmb{q},\tau).
\end{equation}
One can easily solve this equation for $\phi$ in Fourier space, and finally compute the gradient to get the particle displacements through \Cref{eq:irr}. The assumption of irrotationality has been numerically investigated by \cite{Chan2014}, who proved that the curl-free displacement field is an excellent approximation to the full $\pmb{\Psi}$ up to scales $k\sim 1$ \mpch\ even at $z\sim 0$.

To investigate density field statistics, we estimate the density field on a Eulerian cubic grid of $256^3$ cells by the means of a cloud-in-cell mass assignment scheme. We correct for the mass assignment by dividing our density estimates by the cloud-in-cell related kernel \citep{Hockney}. We can neglect the effects of aliasing, as the regime of validity of our results is well below the Nyquist frequency.

Among the statistics of the density field that we consider fundamental and that we examine in this work, the first is the PDF of the density field. The PDF is easily measured by counting the number of cells on the cubic grid with density $[\delta,\delta+\Delta \delta]$ \citep{Klypin17}
\begin{equation}
    P(\delta) = \frac{N_{count}(\delta)}{N_{cell}^3 \Delta \delta},
\end{equation}
normalized by the bin width. We note how the PDF of the density field, often neglected over the years, has recently gained renewed attention (e.g. \cite{Uhlemann16,Repp18,Tosone20}).

Another fundamental statistics we analyze is the power spectrum $P(k)$ of the density field, defined as
\begin{equation}\label{eq:Pk}
    \langle\delta(\pmb{k})\delta(\pmb{k}')\rangle \equiv (2 \pi)^3 \delta^{(D)}(\pmb{k}+\pmb{k}') P (k),
\end{equation}
which quantifies two-point clustering. 

Finally, we also examine the \textit{similarity} between two density fields \textit{A}  and \textit{B}, which we quantify through the cross-correlation 
\begin{equation}\label{eq:cross}
	X(k) = \frac{ \langle \delta_{A}(k)\delta^*_{B}(k) \rangle }{\sqrt{P_A(k) P_B(k)}}.
\end{equation}
While it is always possible to generate two fields that have the same power spectrum, they can differ at higher orders, when evolved: the cross correlation will quantify this by measuring the phases as well. For each approximation we compute the cross correlation the resulting density field with the density field from the $N$-body.

\section{Limitations of Lagrangian Approximations}\label{sect:limits}
\subsection{Truncation}\label{subsect:truncation}
In order to show the limitations of current semi-analytical Lagrangian techniques, we start by considering the truncation approach to model the displacement field. This approach was proposed by \cite{Coles93}, and consists in cutting off the small scale linear power spectrum, just like in \Cref{eq:truncate}, with $W(k)$ chosen as a Heaviside step function. The truncated Lagrangian density field is employed as usual in ZA, namely $\psi_{\text{TZA}}(\pmb{q})=-\delta_\ell(\pmb{q},R)$, where $\delta_\ell(R)$ is the linear density field truncated at a scale $R$. As a result, the real-space cross-correlation coefficient between the approximated Eulerian density field and the Eulerian $N$-body density field is improved with respect to non-truncated schemes. Truncating the Lagrangian density field decreases its amplitude, bringing it closer to the perturbative regime. The cross-correlation, sensitive to phases, is rather insensitive to truncation, which we suspect is why T2LPT excels in this statistic. This motivates us below (\Cref{eq:AT2LPT}), when mixing 2LPT on large scales with a non-perturbative prescription on small scales, to smooth the initial density before computing the 2LPT displacement (unlike ALPT, which smooths the full 2LPT displacement after it is computed). We do not find a substantial quantitative improvement in this switch of the order of operations by itself, but prefer it conceptually.

In the original implementation, a sharp cutoff for the window function $W(k)$ was adopted. \cite{Melott94} obtained a further improvement by adopting a Gaussian smoothing window. Ideally, the scale of cutoff should be the one that marks the onset of nonlinearities, when the variance of the density field is $\sigma^2 \sim 1$. While the cross correlation improves, the power spectrum between truncated and non-truncated schemes remain similar. This suggests that the improvement is due to the better modelling of the \textit{phases} of the field. Extending the truncation procedure to higher order corrections improves over the TZA \citep{Buchert94,Melott95}.

\begin{figure}
    \centering
    \includegraphics[width=0.45\textwidth]{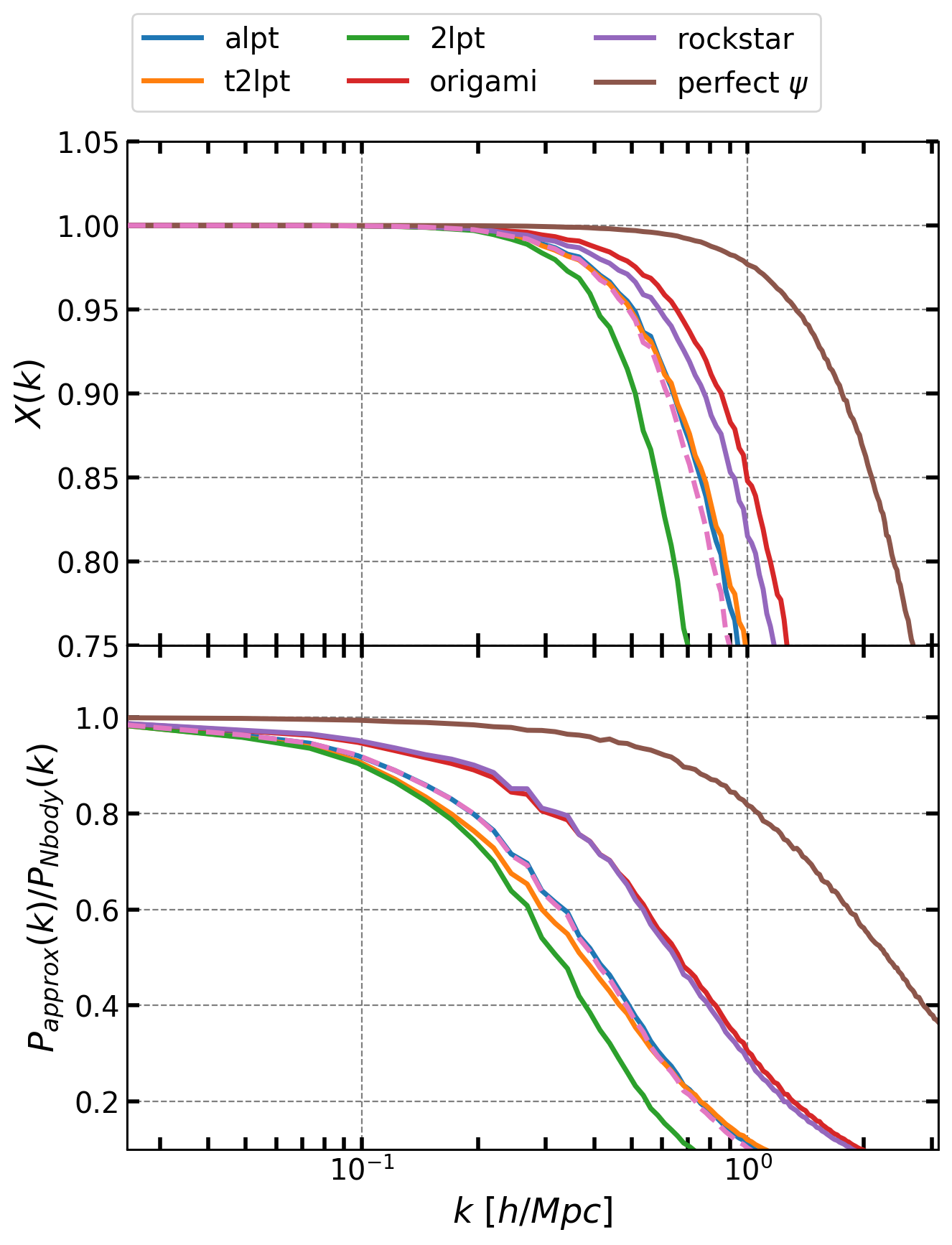}
    \caption{Cross correlation (top) and power spectrum (bottom) of the density field produced by the approximations of the displacement field considered in the text. Surprisingly, T2LPT approach performs as well as ALPT in these statistics, despite its poor modeling of the overdense regions. An attempt to reintroduce the SC information with AT2LPT (the only dashed line), performs almost identically to ALPT and T2LPT. \origami\ and \rockstar\ labels refer to approaches where the displacement field is modelled by \Cref{eq:origami1,eq:origami2}. In these cases the small scale component has been set to $\psi=-3$ in correspondence of halo regions detected by these halo finders on the reference simulation (and so they are not predictive). This suggests that an improvement of Lagrangian approximations schemes should come from a better detection of overdense regions. "Perfect $\psi$" refers to a realization whose density field was generated by the curl-free displacement as measured from the $N$-body.
    \label{fig:Fig1}}
\end{figure}

Following a standard method (see e.g. \cite{Neyrinck13,Kitaura13_ALPT,Munari17}), we compare the power spectrum and the cross correlation of dark matter fields obtained through various approximations in \Cref{fig:Fig1}, where we see that truncated schemes perform well at the level of Fourier-space cross correlation statistics as well, as we show for truncated 2LPT (T2LPT). For the latter we adopted an interpolation scale of $2.5$ \hmpc\ while for ALPT we used the a smoothing scale of $3.0$ \hmpc. These values were chosen as to maximize their cross correlations with the reference simulation, and were easily fine-tuned by a simple inspection of a handful of different values. It is impressive to note how T2LPT yields something comparable to ALPT in the cross-correlation, even though T2LPT truncates all the information on small scales, which give overdense regions that are the first to collapse. However, as shown below in \Cref{fig:Fig8}, T2LPT fails badly at modeling the PDF, and it performs even worse than 2LPT, which already performs poorly in both overdensities and underdensities. This result suggests that the only benefit of the simple SC prescription in \Cref{eq:sc} is an improved description of voids. But as we show in the following, the problem with the simple SC prescription is that it detects collapsed voxels too naively, looking for collapses only at the grid scale.

To test the hypothesis that the SC prescription is of little aid in modeling overdense regions, we reintroduce \Cref{eq:sc} by following the same baseline of ALPT: we augment T2LPT with the SC formula, obtaining an "Augmented Truncated Lagrangian Perturbation Theory" (AT2LPT)
\begin{equation}\label{eq:AT2LPT}
    \psi_{\text{at2lpt}} = \psi_{\text{t2lpt}}(\sigma_R) + \psi_{\text{sc}} \circledast (1- \mathcal{G(\sigma_R)}).
\end{equation}
The large-scale component corresponds to T2LPT Gaussian smoothed at the scale $\sigma_R$, while the small scale component corresponds to the usual SC formula. Notice how, unlike in ALPT, $\psi_{\text{t2lpt}}$ is not convolved with a window function because the convolution has been already applied to the linear density field.

\begin{figure*}
    \centering
    \includegraphics[width=0.8\textwidth]{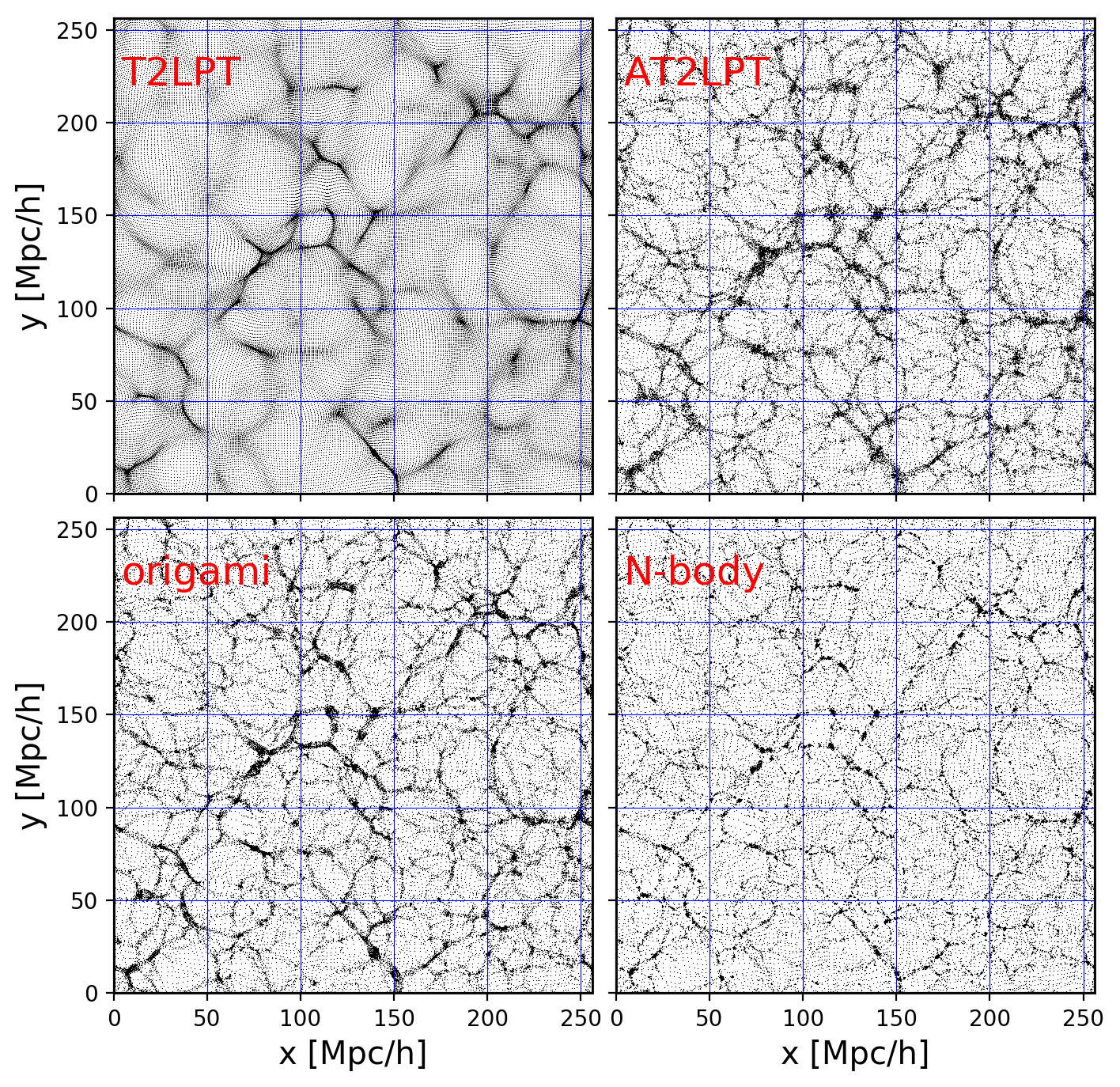}
    \caption{Eulerian positions of particles in a one-particle-thick Lagrangian slice of the simulations. One can notice how T2LPT traces well the cosmic web, but is not able to reproduce collapses or voids. The augmentation of T2LPT with SC (Augmented Truncated LPT) recovers the voids more correctly, but loses the benefit of truncation for filaments and haloes. The approach based on detecting halo particles through \origami\ is the one that looks more similar to the $N$-body.}
    \label{fig:Fig2}
\end{figure*}

One expects that including SC should improve the power spectrum or the cross correlation statistics. What we show in \Cref{fig:Fig1} is that this is not the case, and that AT2LPT is in fact comparable to T2LPT. This confirms that the only benefit of \Cref{eq:sc} is to be an extremely good description of the evolution of voids, but we find that it does not perform as well in overdense regions. SC does not collapse structures in the same way as the $N$-body does, as shown in \Cref{fig:Fig2}. We can see that the introduction of the SC formula expands haloes and filaments with respect to T2LPT.

\section{Haloes in the Lagrangian Picture}\label{sect:halofinders}
Based on the results of truncated schemes, we have thus realized that a shortcoming of Lagrangian approaches is a poor modelling of overdense regions.

First, we investigate whether tidal-field-related quantities could provide additional information to model the displacement field. One motivation for investigating the tidal field is the success of the SC approach for void particles; this suggests that the `separate universe' approximation to uncollapsed cells is highly accurate, even for anisotropic cells. \citet{DaiEtal2015} found that corrections to a patch's evolution are largely contained in the tidal field. Another motivation is that one can think of the eigenvalues of the tidal field as marking the onset of shell crossing, so they could be able to trace more closely halo formation and overdensities. While we find several correlations, none of them seem to be able to improve the modelling of $\psi$, with respect to the standard linear density field. We report these results in \Cref{appendix:tidal}.

In the following discussion we then focus on haloes. The advantage is that we already know an approximation to model halo voxels in terms of the Lagrangian displacement, i.e. just setting $\psi=-3$ \citep{Neyrinck13}. Our guess is that a more accurate detection of halo patches from the initial conditions yields an improvement in the statistics of the resulting density field. We can formally prove this by exploiting the \textit{exact} information about halo regions, extracted from the reference simulation through halo finders.

\subsection{Origami-informed realization}
\origami\ is a Lagrangian halo finder which tracks the trajectories of the particles in the simulation to detect shell crossing events. Whenever trajectories overlap along a principal axis, one can consider that particles collapsed along that direction. The number and the direction of the collapses is indicative of the morphology of a particle.

As anticipated, we use \origami\ \citep{Falck12} to detect halo particles in our reference simulation. We exploit this \textit{a posteriori} information from the $N$-body result to realize a simulation whose displacement field has the same functional form as  \Cref{eq:AT2LPT}:
\begin{equation}\label{eq:origami1}
    \psi_{\text{origami}} = \psi_{\text{t2lpt}}(\sigma_R) + \psi_{\text{ss}}\circledast (1- \mathcal{G(\sigma_R)}),
\end{equation}
but this time the small scale component $\psi_{\text{ss}}$ is
\begin{equation}\label{eq:origami2}
\psi_{\text{ss}}=
\begin{cases}
-3, &\text{\origami\ halo regions},\\
\psi_{\text{sc}} &\text{everywhere else},
\end{cases}
\end{equation}
namely it is set to $\psi=-3$ for halo voxels as found by \origami. All the non-halo regions are modeled with the SC formula, while T2LPT contributes to both halo and non halo regions on large scales.

In \Cref{fig:Fig1} we show the cross correlation and power spectrum statistics for this approach. We can see the noticeable improvement over ALPT, T2LPT and AT2LPT. This proves our point that the detection of halo regions alone could be the determining factor in modelling the displacement field. This improvement can also be seen in terms of particle displacements in \Cref{fig:Fig2}: the displacement based on halo particles found through \origami\ is the closest to the actual $N$-body result. In \Cref{fig:Fig1} we also show the result of the curl-free displacement field, measured directly from the simulation, dubbed \textit{perfect $\psi$}. This indicates the limit that one could hope to achieve through a scalar displacement field $\psi(\pmb{q})$ that satisfies \Cref{eq:irr}. As we can see, even adding back the exact halo information does not saturate the full information of the exact irrotational displacement.
\begin{figure*}
    \centering
    \includegraphics[width=0.75\textwidth]{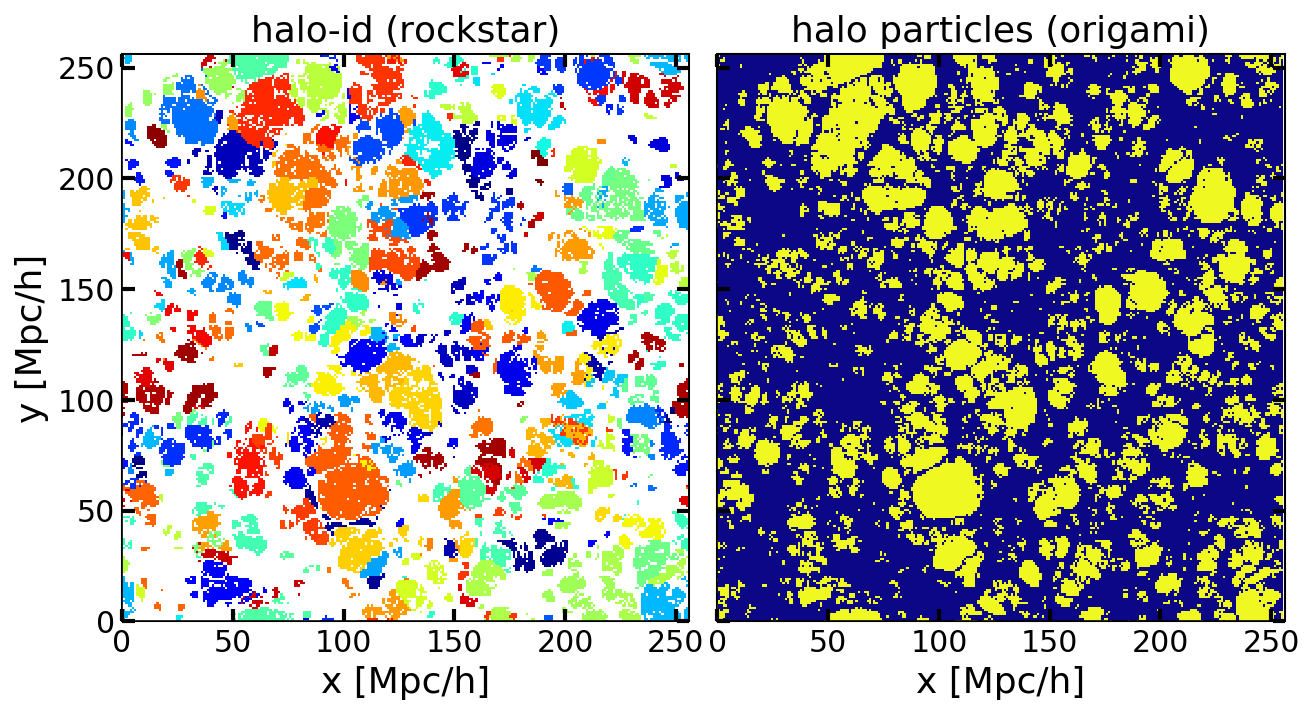}
    \caption{$1$ \hmpc-thick Lagrangian slice of the initial conditions. On the left we show halo particles color coded randomly based on their halo-id, identified by \rockstar. On the right we show halo particles found by \origami. Their agreement is rather high, considering the difference in their baseline process.}
    \label{fig:Fig3}
\end{figure*}

\subsection{Rockstar-informed realization}
The way \origami\ detects haloes does not necessarily provide the same results as for the more standard Eulerian-based definition. We repeat the same procedure with halo particles detected by \rockstar\ \citep{Behroozi13}. We find a comparable result to \origami\ (\Cref{fig:Fig1}), which is surprising if we consider the difference in their baseline procedure to detect haloes. The similarity in the halo particles detected is also shown by \Cref{fig:Fig3}.

For halo particles detected by \origami\ and \rockstar, in \Cref{eq:origami2} we also tried to substitute the actual value of $\psi$ extracted from the reference $N$-body, finding virtually no change in $X(k)$ and $P(k)$ over the case of setting them to $\psi=-3$. This result further stresses that the detection of halo particles is the main problem, and modelling them as $\psi=-3$ suffices.
\begin{figure*}
    \centering
    \includegraphics[width=0.8\textwidth]{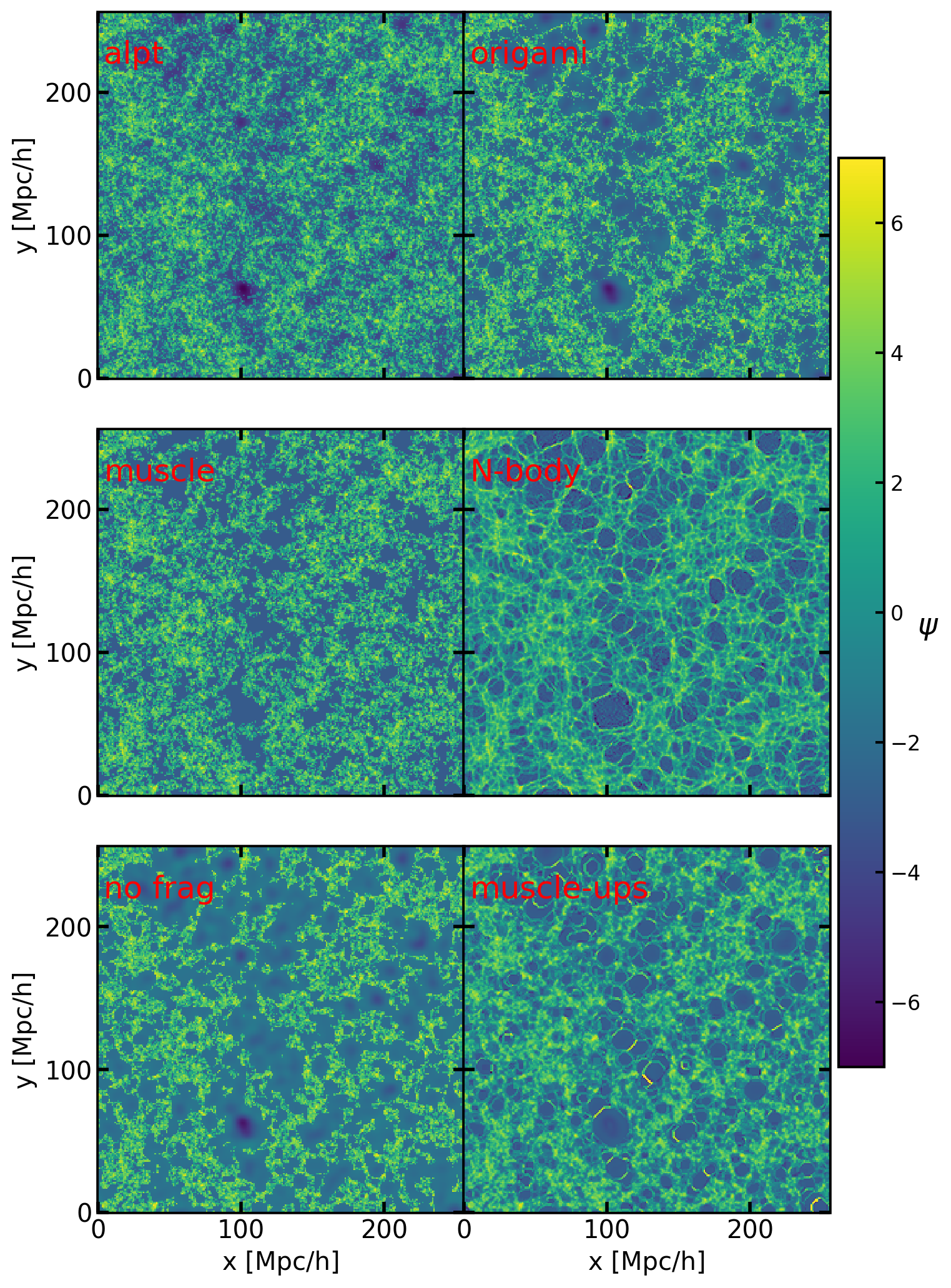}
    \caption{One-particle-thick Lagrangian slice of the displacement field of the various approximation schemes. Particles have been colour-coded according to the value of $\psi$. ALPT and \muscle\ can trace well the innermost halo voxels when compared to \origami\ and $N$-body. Our method of expanding them further according to \Cref{eq:tw3} is shown in the bottom left panel. It increases the similarity to the $N$-body, with the limit that there is no fragmentation between the halo patches. In \Cref{sect:toyhalo} we explain how the fragmentation can be reintroduced with the addition of a halo model, as shown in the bottom-right panel.}
    \label{fig:Fig4}
\end{figure*}

\section{MUltiscale Spherical Collapse Lagrangian Evolution Using Press-Schechter}\label{sect:muscleups}
In this section we outline the algorithm we propose to predict more accurately which particles collapse into haloes
from the initial conditions.

\subsection{Extended Press and Schechter Formalism (EPS)}
In the following we refer to haloes in a Lagrangian sense; more correctly we refer to proto-halos that can be detected through \origami\, and they do not necessarily correspond to haloes in the more common Eulerian sense and as they are found in standard halo finders. More on this is discussed at the end of \Cref{sect:toyhalo}.

In \Cref{fig:Fig4} we show a slice of the scalar displacement field in various approximations. Haloes are dark patches with $\psi \sim -3$. It is clear that ALPT and \muscle\ cannot account for haloes at a level comparable to the $N$-body, but they seem to detect well the innermost regions of the halo patches as measured by \origami. This suggests that one could try to \textit{expand these regions} to increase the resemblance with the $N$-body case.

Before proceeding, here we recall the EPS formalism: if the linear density field smoothed at a scale $R$ is above a threshold at a point $\pmb{x}$ in the initial conditions, i.e. $\delta_\ell(\pmb{q},R)>\delta_c/D_1$, all the particles that are within a distance $R$ from $\pmb{q}$ form a halo of mass $M>4\pi \rho_m R^3/3$ \citep{Press74}, where $\rho_m$ is the average matter density of the Universe. The EPS formalism is not exploited by semi-analytical simulation schemes. The only similar process can be found in \muscle, since halo regions are detected through the multi-scale smoothing of the linear density field, but \muscle\ is ultimately voxel-based as the condition $\psi=-3$ is set independently for each voxel. 

The equivalent of the EPS formalism in terms of the displacement field would be to set $\psi=-3$ for all the voxels within the patch identified by the smoothing filter. Intuitively, it is reasonable to consider that a voxel does not undergo a standalone collapse, especially if the collapse criterion is satisfied at scales larger than the inter-particle separation. Also, it is already known that the standard SC criterion may succeed to detect the peaks but fail to capture the patch \citep{Ludlow11}. However, it is not necessarily true that there is a one to one correspondence between the smoothing scale of the window function and the size of the proto-halo patches measured in $N$-body, and this depends also on the choice of the filter (see \cite{Chan15}). Yet, the fact that these patches are roundish suggests that there may be a proportionality to relate smoothing scale and size for some appropriate choice of the filter function.

In the original EPS implementation \citep{Bond91}, the mass of haloes is set by a top-hat window function
\begin{equation}\label{eq:TopHatWindow}
    W(k) = 3\frac{\sin{(kR)}-kR\cos{(kR)}}{(kR)^3},
\end{equation}
because the smoothing scale $R$ can be identified with the radius of a proto-halo of size $R$, based on the SC toy model. The correct correspondence becomes less obvious for general window functions like a Gaussian one
\begin{equation}\label{eq:GaussWindow}
    W(k) = e^{-k^2 R^2/2}.
\end{equation}
For example, it was already noted by \cite{Bond96} that the mass enclosed in a proto-halo of size $R$ identified through a Gaussian filter should correspond to a proto-halo size of about $R/2$ with a top-hat filter, if we impose that they enclose about the same mass.

\begin{figure}
    \centering
    \includegraphics[width=0.45\textwidth]{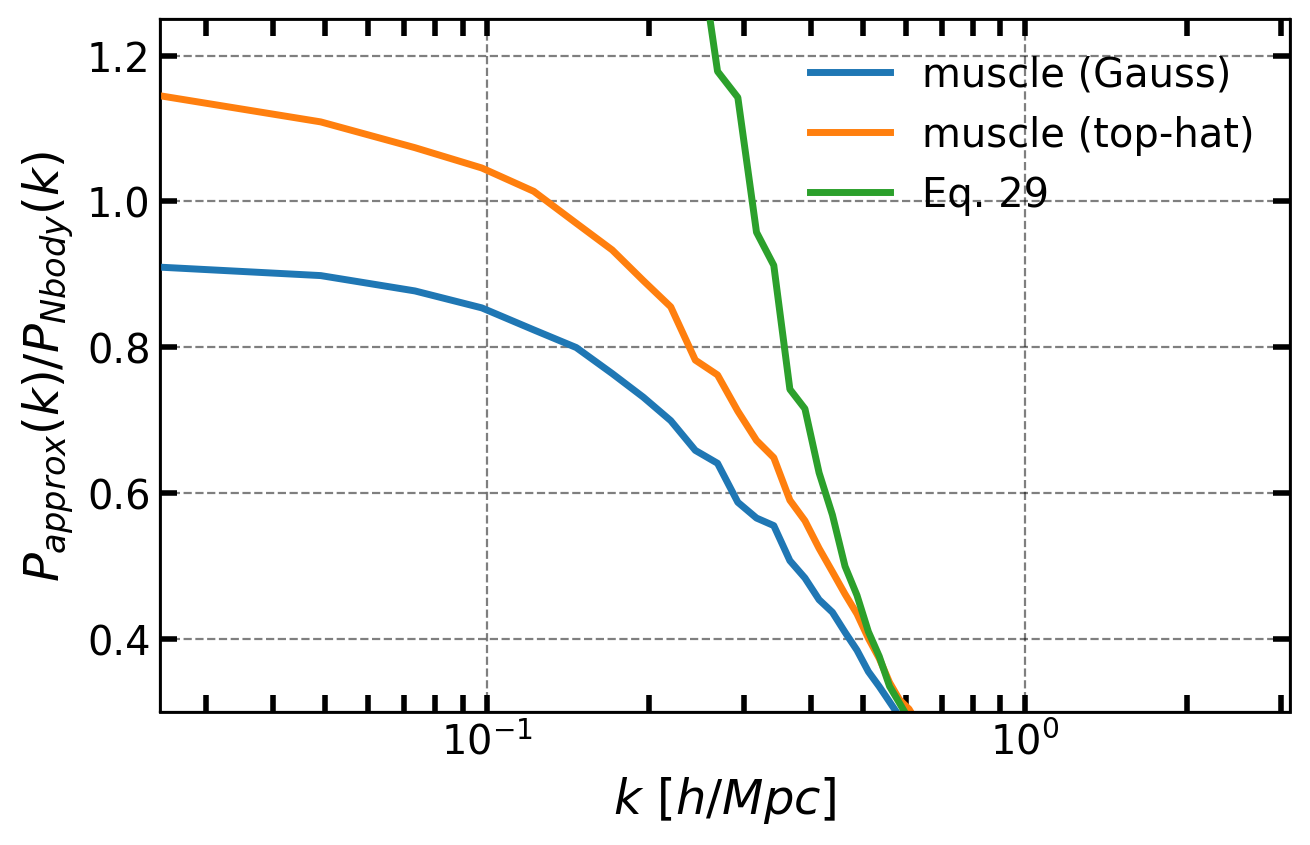}
    \caption{Difference in the power spectrum of a density field realized through \muscle\ with Gaussian and top-hat filters. The power is directly affected by the number of particles that satisfies the SC criterion, a number that increases with the top-hat filter with respect to the Gaussian filter. When the density field is generated through \Cref{eq:tw3} by setting $\psi=-3$ around collapsing voxels, that is by expanding the proto-halo patches, there is a further increase of power. This offset can be correct by implementing an interpolation in order to match the large-scale power to perturbation theory predictions, as prescribed in \Cref{eq:tw1}.}
    \label{fig:Fig5}
\end{figure}

To illustrate the impact of the choice of different filters, in \Cref{fig:Fig5} we compare the power spectra obtained with \muscle\ by adopting both a top-hat and Gaussian window function for detecting halo particles. The original implementation of \muscle\ makes use of a Gaussian, and slightly underestimates the power spectrum, while a top-hat based \muscle\ implementation generally has a slightly larger amplitude than the reference power spectrum. This difference comes from the number of halo particles detected: in the top-hat case the condition for collapse is satisfied to a maximum scale that can be as large as twice the maximum scale of collapse of the Gaussian window, thus detecting more halo particles. These are responsible for the offset of the power spectrum even on linear scales.

\subsection{\muscleups}
Among all the possible implementations, we opt for an EPS scheme with a Gaussian window: all the particles that are within a radius $R$ from a voxel where the condition of SC is satisfied are set to $\psi=-3$. With this choice there is a close correspondence with the halo patches found through \origami, as shown in \Cref{fig:Fig4}. We chose a Gaussian for its simplicity, and because it seems to work well with a factor of unity between the smoothing length and the radius of the sphere assumed to collapse. Other choices might further improve results, e.g.\ a different such factor, and a different window shape, but we do not explore this parameter space, already finding a simple solution.

This approach has the counter-effect of increasing substantially the number of halo particles, and it does introduce the problem that the amplitude of the power spectrum overshoots the amplitude of the target spectrum at linear scales, even though we are using a Gaussian window (\Cref{fig:Fig5}). Luckily, we already know how to circumvent this limitation, i.e.\ by interpolating the large scales, where we can use T2LPT, with the small scales, where now we have a better description for halo particles. This is the same scheme we used in the case halo particles were detected through \origami\ and \rockstar\ in \Cref{eq:origami1}, i.e.
\begin{equation}\label{eq:tw1}
    \psi_{\text{muscleups}} = \psi_{\text{t2lpt}}(\sigma_R) +\psi_{\text{eps}} \circledast (1- \mathcal{G}(\sigma_R)),
\end{equation}
but this time the small scale term \Cref{eq:origami2} is substituted by a fully analytical expression: whenever at some time $\tau$ the smoothed linear density field at some time $\delta_\ell(\pmb{q},R) D_1(\tau)$ (defined in \Cref{eq:truncate}) exceeds the threshold of collapse $\gamma=1.5$ at any scale $R$, including the trivial case of the inter-particle distance of the simulation, we should set $\psi=-3$ for all neighbouring voxels within a distance $R$ from the voxel $\pmb{q}$ under examination. Formally it reads
\begin{equation}\label{eq:tw3}
\psi_{\text{eps}}(\pmb{q})=
\begin{cases}
-3, &\text{if}~ \delta_\ell(\pmb{q},R)>\gamma/D_1,\\&\text{for any}~\pmb{y}\in \mathcal{B}_R(\pmb{q}),\\&\text{and}~R\ge 0;\\
3\left[ \left(1-D_1(\tau)\frac{\delta_\ell(\pmb{q})}{\gamma} \right)^{\gamma/3} -1 \right]  &\text{otherwise}.
\end{cases}
\end{equation}
Here we used the notation
\begin{equation}
    \mathcal{B}_R(\pmb{q}) = \{ \pmb{y} \text{ such that } \Vert \pmb{q}-\pmb{y}\Vert \leq R \}
\end{equation}
to refer to the `ball' of points which are at most at a distance $R$ away from $\pmb{q}$. 

\Cref{eq:tw1} prescribes a combination of T2LPT truncated at the scale $\sigma_R$ for the large-scale component, \textit{augmented} with the small scale component \Cref{eq:tw3}, which corresponds to our adaptation of the EPS formalism to a displacement field. If the condition of collapse is never met, the equation in the second line of \Cref{eq:tw3} is used, which is the SC formula. Even if a voxel does not meet the condition for collapse, it could still have $\psi=-3$ if it is near a collapsing voxel. 

We dub the scheme to generate this displacement field MUltiscale Spherical Collapse Lagrangian Evolution Using Press-Schechter (\muscleups). Admittedly, $\psi_{\text{eps}}$ should be called $\psi_{\text{muscleups}}$, since this part would coincide with \muscle\ if $\mathcal{B}_{R=0}(\pmb{q})=\{\pmb{q}\}$, namely if we removed the voxel-expansion process. To avoid confusion due to many names, we decided to refer to the whole displacement \Cref{eq:tw1} as \muscleups, also because the results from \Cref{eq:tw3} alone overestimate the power on large scales (see \Cref{fig:Fig5}).

As in \muscle, we perform the smoothing process by starting from the inter-particle distance, and progressively moving to larger distances; this process stops as soon as a scale where no voxel collapses is reached. Also, we increase the number of smoothing filters to cover all possible discrete lengths in units of inter-particle distance. Just like in \muscle\ and in SC, the process of setting all voxels to the same value introduces a nonzero average $\langle \psi \rangle= C$ that must be subtracted to preserve the zero mean density condition.

To conclude, we want to stress that \Cref{eq:tw1} is only the displacement field part of the complete \muscleups\ algorithm. This displacement is plotted in \Cref{fig:Fig4} with the label \textit{no frag}, where one can see that proto-halo patches are not fragmented into separate haloes as well as in $N$-body. We think we can do better if we were able to separate them in a way that mimics the formation of haloes. In the following section, we complete \muscleups\ by addressing this issue.

\section{Building a Halo Catalogue and the Halo Model}\label{sect:toyhalo}
The natural follow-up to a multi-scale smoothing of the initial density is to build a halo catalogue. In this way \muscleups\ can output a displacement for CDM particles and for the haloes as well, which is more useful for the purpose of mock galaxy catalogues.

There are already EPS based algorithms available, such as \peakpatch\ \citep{Bond96} and \textsc{pinocchio} \citep{Monaco01}. In this section we show how we can build a halo catalogue as well, with a simpler approach because it does not require to make assumption about the gravitational dynamics, but capable to achieve higher precision for the HMF. Finally, we will also implement a halo model in order to mimic the collapse of the particles of the of proto-haloes.

\subsection{Building a halo catalogue}
The main idea of the EPS formalism is that haloes form from the collapse of spherical patches, overdense with respect to the background density. The cloud-in-cloud problem of the original argument by Press and Schechter \citep{Press74} was addressed in the EPS approach by considering the largest patches that collapse by the redshift under examination \citep{Bond91,Bond96}. This is consistent with a hierarchical structure formation picture, where a large object formed by processes of mergings of smaller halos or by accretion, thus subsuming the smaller overdensities.

For each particle, we save the largest smoothing scale at which the smoothed density field \Cref{eq:truncate} collapses, i.e. $\delta_{\ell}(\pmb{q},R) \geq \gamma/D_1$, if it collapses at all. We dub the array containing this information the \textit{sift field} $\mathcal{S}(\pmb{q},R)$; in analogy to sifting grains, it stores the hole size $R$ at which a particle in Lagrangian space falls through the sifter. This can be stored in terms of units of the inter-particle distance, thus as an integer. 

The sift field alone does not suffice to build a catalogue, as we have a discrete number of smoothing filters and many candidate haloes which share the same smoothing scale. For the sake of ordering them by the time of their collapse at a fixed smoothing scale, we also store the \textit{collapse condition field} $\mathcal{CC}(\pmb{q},R,\tau) = 1-D_1(\tau) \delta_\ell(\pmb{q},R)/\gamma$, evaluated at the value $R$ of the sift field $\mathcal{S}(\pmb{q},R)$, for each collapsing particle in Lagrangian space. Here $\gamma = 1.5$ is the value adopted in the SC formula \Cref{eq:sc}.

It is worth noticing that if the smoothed linear overdensity were continuous over $R$, it would bypass the need of the $\mathcal{CC}$ field, as the time of collapse would be implicitly contained in the smoothing radius itself.

It suffices to create $\mathcal{S}$ and $\mathcal{CC}$ arrays at the beginning of the bottom up filtering process of \muscleups, so that for each particle we can save or update the largest smoothing scale at which the collapse condition is verified, stored in $\mathcal{S}$, alongside the corresponding density stored in $\mathcal{CC}$. At the end of the filtering process, we have a list of all the halo particles which satisfy the collapse condition at some scale, stored in $\mathcal{S}(\pmb{q},R)$. These particles are all potential \textit{seeds} of proto-haloes, since they can be associated to patches of size $R$. We think that the name \textit{seeds} is more apt to refer to these particles, as they correspond to the \textit{centers} of the proto-halo patches in the initial conditions, but as it will be clear in the following the final shape of a patch may not be spherical.

To build a halo catalogue, we first sort the halo seeds saved in $\mathcal{S}(\pmb{q},R)$ based on their smoothing scale, from the largest the smallest. Then, particles with the same smoothing scale $R$ are sorted in decreasing order of smoothed linear density, namely in order of increasing $\mathcal{CC}$ values. Since each of these particles can be regarded as the seed of a patch, this procedure yields a list of all the candidate proto-halo patches ordered based on their size, and for each size they are ordered based on their densities as well. 

At a fixed smoothing scale $R$, we progressively consider particles from the densest to the least dense in the list. For each halo seed, we cycle through all the particles within a distance $R$ from it in Lagrangian space. The same unique halo-id is assigned to each of these surrounding particles within the radius only if (1) the particle has $\psi=-3$ assigned, namely it is considered a halo particle according to \Cref{eq:tw3}, and (2) the particle has not been assigned a unique halo-id. A particle may have a halo-id already assigned because it lies within a larger or a denser patch than the one under examination.

\begin{figure}
    \centering
    \includegraphics[width=0.47\textwidth]{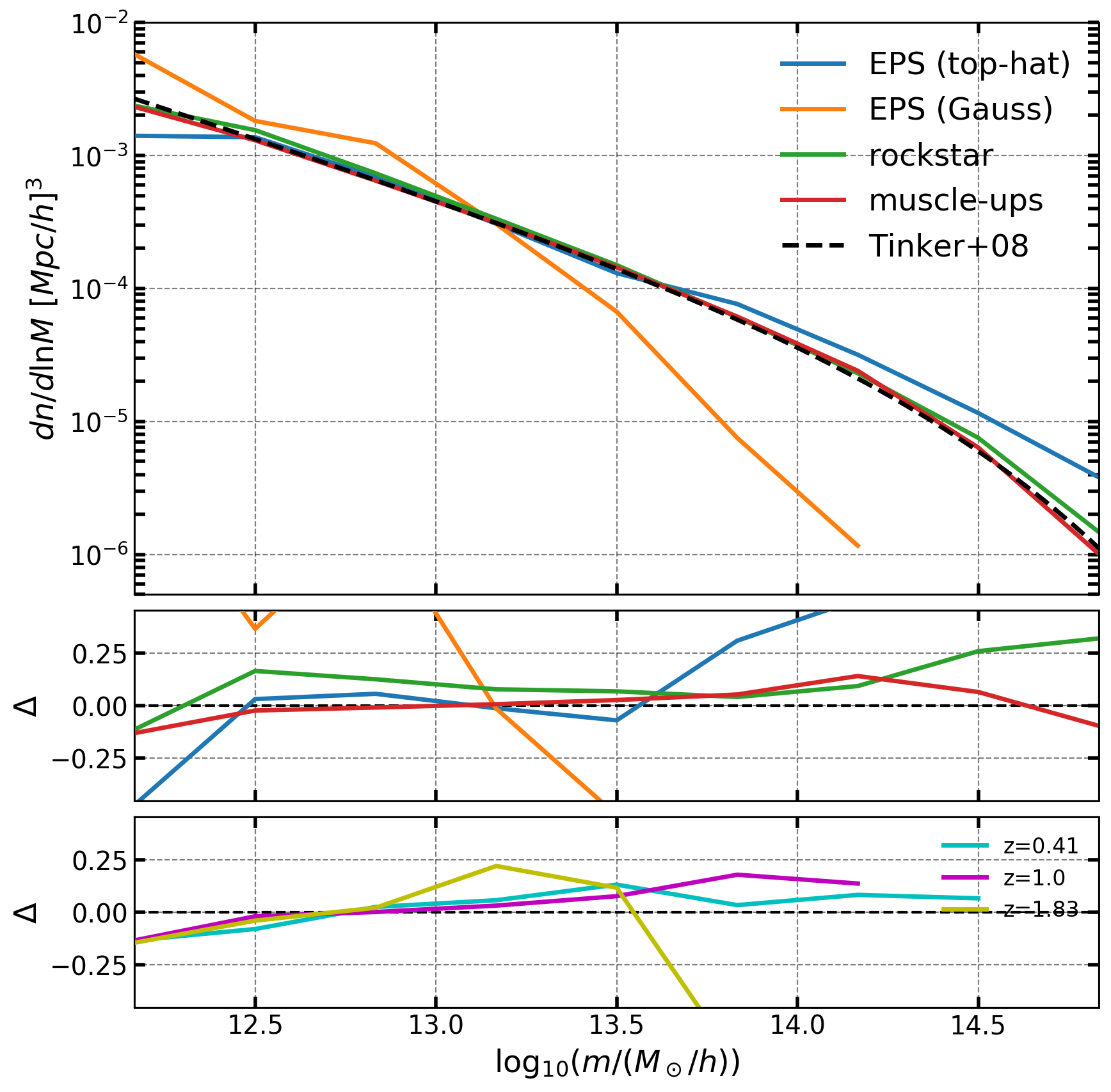}
    \caption{In the top panel we show the HMFs obtained in various scenarios at $z=0$. \textit{EPS (Gauss)} refers to an EPS approach with a Gaussian window function, when no percolation occurs. Analogously for the \textit{EPS (top-hat)} result. In \muscleups\ we implement the scheme discussed in the text based on percolating Gaussian proto-haloes to match a target HMF. The \rockstar\ label corresponds to the HMF measured directly from the reference simulation.
    In the middle panel we show the relative residuals of the lines in the top panel with respect to the fit provided by \protect\cite{Tinker08}, which we chose as the target HMF and whose implementation is found in \textsc{colossus} \protect\citep{Diemer17}. In the bottom panel, we show the relative residuals of \muscleups\ HMFs at higher redshifts with respect to the corresponding target fits of \protect\cite{Tinker08}.}\label{fig:Fig6}
\end{figure}

As we examine the particles in the sift field, it may occur that a halo seed has been assigned already assigned a \textit{parent} halo-id. In this case there are a couple of ways to proceed: (a) either we discard it, or (b) we \textit{merge} its patch into the parent halo. As in the original work of \cite{Bond96}, the  term `merging' here might be misleading, as there is not necessarily an actual, dynamical merging process occurring between the two haloes in our semi-analytical approach. Effectively, combining proto-halo patches based on their overlap is a \textit{percolation} rather than a merging, but the term merging conveys the idea that in a real $N$-body simulation, two overlapping patches should really merge. For convenience, in the following we use both terms interchangeably.
In \Cref{fig:Fig6} we present the HMF resulting from the first scenario, where no percolation occurs because a hard exclusion criterion is applied. We see that it drastically fails when compared to both the theoretical prediction of \cite{Tinker08} and the HMF measured from our reference simulation with \rockstar. For both \rockstar\ halo finder and for the reference HMF of \cite{Tinker08}, the spherical overdensity $\Delta$ is set by \Cref{eq:B&N}, which we will also adopt in our implementation of the halo model. We see that a Gaussian-based EPS results in both an excess of small haloes and in a complete lack of large haloes, which suggests that percolation is necessary when a Gaussian filter is adopted. We also tried to change the proportionality scale between the smoothing scale $R$ of the Gaussian filter and the corresponding size in Lagrangian space of the corresponding proto-halo, to see whether there was a way to recover the correct HMF, but this only results in shifting the mass scale at which an excess of halos occur, based on the value of the proportionality scale.

On the other hand, in the case of a top-hat-based EPS, more haloes are found at higher masses, and the resulting HMF seems a better approximation to the expected HMF, even without percolation. However, its predictions are not quantitatively reliable, and we would like to develop an algorithm that yields a HMF closer to the expected result; the Gaussian filter allows more freedom to do this.

Attempting to predict the formation and the percolation of haloes from the initial conditions at the level of each individual object is a daunting task. \peakpatch\ and \textsc{pinocchio} are the only software that attempted this, with some success. Here we propose a different scheme that exploits the excess of small haloes candidates detected through Gaussian smoothing; the great deal of freedom in merging these halo candidates allows us to build a halo catalogue that can reproduce the HMF measured from $N$-body simulations. The idea is to insert a halo candidate in the catalogue or to perform a merging between two haloes if and only if this improves the resulting HMF by making it closer to the target. This means that we need to update the resulting HMF on the fly for every merging or whenever a new candidate halo is inserted in the catalogue. But this process is cheap considering that we already have a list of all the possible haloes.

The result of this process is plotted in \Cref{fig:Fig6}, where we can see that it minimizes the residuals with respect to the target HMF, being extremely precise over a large mass range. It is worth stating explicitly that this optimization procedure of the HMF is not stochastic, but it is fully determined from the initial conditions once the internal parameters of the algorithm have been specified. These internal parameters include the number of smoothing scales in the initial conditions, that we fixed to cover all possible discrete lengths in units of inter-particle distance, and the sampling of the target HMF. The sampling has only one free parameter, the width of the mass bin $\Delta \log(M/(M_{\odot}/h))$. In fact, the mass of the smallest halo is implicitly fixed by the mass resolution of the simulation, while the largest expected halo has a mass that can be easily computed from the target HMF, knowing the box-size of the simulation. We tried several resolutions for the width of the mass bins, finding little variation for a high number of bins. The results we present here were obtained with a width of $\Delta \log(M/(M_{\odot}/h))=0.025$.

There is a limitation in our result: by matching exactly a template HMF, we cannot reproduce the intrinsic fluctuations that exist in the HMF measured in the reference $N$-body with the same initial conditions. In other words, the variance of the measured HMF from many realizations derived with our scheme, is expected to be smaller than the corresponding variance from $N$-body realizations. We do not have a solution for this, but arguably this problem also affects, in a different way, the other semi-analytical schemes that try to reproduce the HMF from the initial conditions: the fluctuations of the HMF estimated from the initial conditions do not coincide with those measured in the Eulerian HMF at low redshift.

As a first final thought, we would like to underline that the fragmentation process is, of course, not exact. Even if we could identify with perfect accuracy the halo particles in the initial conditions, it is still not obvious how to group them into halo patches. \Cref{fig:Fig3} shows the similarity of the halo particles detected between \origami\ and \rockstar, but there is not a one-to-one correspondence between an \origami\ isolated patch and the halo-id as found by \rockstar. We cannot build a Eulerian HMF from the \origami\ morphology alone, and some choices based on the density field must be made \citep{Falck12}. The HMFs from \origami\ and \rockstar, in fact, are found to be different; by detecting only particle crossings and not testing for gravitational boundedness, \origami\ tends to identify larger patches than \rockstar~\citep{Knebe11}.

Here we want to highlight the differences of \muscleups\ with respect to \peakpatch\ and \textsc{pinocchio}. \peakpatch\ first detects overdense patches in the field through a top-hat window function; afterwards, haloes are built around the densest voxel in each patch by including spherical shells. These halo patches are evolved through homogeneous ellipsoidal collapse dynamics, to determine the size of the patch that collapses under the redshift under examination, when the overdensity $\Delta$ of the halo is below a certain threshold.
\textsc{pinocchio}, on the other hand, solves the equations of ellipsoidal collapse for every particle to determine the time of collapse. Collapsing particles are grouped into haloes afterwards, based on their distances in Eulerian space after they have been displaced through ZA.

On the other hand \muscleups\ works only at the level of the patches detected through a Gaussian window, and does not assume the gravitational dynamics of collapse to determine the final haloes. While the regions where $\psi=-3$ are round, due to the voxel expansion process of \Cref{eq:tw3}, the way patches are fragmented into haloes does not impose any spherical symmetry, and patches can be aspherical. Due to the fact that Gaussian proto-patches are much smaller than their top-hat counterparts, we can percolate them to match a target HMF, instead of trying to get the HMF directly from the top-hat proto-patches. 

It is worth mentioning that the CUSP formalism \citep{Manrique95,Manrique96} adopts a closely related idea, based on the ansatz that virialized objects can form through subsequent merging and accretion events of non-nested peaks detected through Gaussian smoothing, and they modified accordingly the peak model framework of \cite{Bardeen86} with new definitions for the critical overdensity and halo mass. In this regard, we share the same idea of percolating the halo candidates, so that we can optimize the HMF, but we do not employ their formalism or their prediction for the halo merger tree; we regard our approach closer to the EPS framework, as it is based on overdense patches rather than peaks.

To conclude, unlike \peakpatch\ and \textsc{pinocchio}, we do not need to assume the dynamics of collapse, as we can evolve the particles first according to the displacement of \Cref{eq:tw1}, and then displace them in Eulerian space to match the expected halo density profile. This is explained in the following.

\subsection{Implementing the Halo Model}
The working hypothesis of the halo model is that all the matter in the Universe is found in virialized haloes, although this is a simplification \citep{Angulo10}. The halo model power spectrum of CDM can be written as the combination of two contributions: the large-scale correlations between haloes, commonly referred as the two-halo term, and the correlation between particles inside the same halo, the one-halo term, \citep{Peacock00,Seljak00}
\begin{equation}\label{eq:halomodel}
P(k) = P_{2h}(k) + P_{1h}(k).
\end{equation}

Schemes like \peakpatch\ and \textsc{pinocchio} use this assumption, as they first determine proto-haloes in the initial conditions, which are then displaced according to perturbation theory. Implicitly this corresponds to a displacement field of the form
\begin{equation}\label{eq:psihalomodel}
    \psi = \psi_{2h} + \psi_{1h}.    
\end{equation}

We prefer to follow an alternative approach to this problem: since we already created a halo catalogue from the initial conditions, one could redistribute particles that have been tagged with the same halo-id into a Navarro Frenk and White (NFW) density profile \citep{Navarro96}, directly in Eulerian space. This avoids assuming \Cref{eq:psihalomodel}, namely that the two-halo displacement and the one-halo displacement are not correlated. We can first displace particles according to a Lagrangian scheme of our choice, and then we redistribute them into NFW haloes, as we explain in \Cref{appendix:halo_displacement}.

This scheme recalls the \textsc{PThalos} procedure in redistributing particles directly in Eulerian space \citep{Scoccimarro02,Manera13}. The difference is that \textsc{PThalos} is a fundamentally Eulerian approach: after the 2LPT displacement of particles is run, a halo finder is employed to detect the densest regions (with a very large linking length, necessary because 2LPT haloes are far overdispersed). These are fragmented to match the desired HMF, and their particles are sampled from NFW profiles. In contrast, in order to match the correct HMF, our fragmentation is implemented in the initial conditions, during the compilation of the halo catalogue.

\section{Density field statistics from \muscleups}\label{sect:results}
\begin{figure*}
    \centering
    \includegraphics[width=0.75\textwidth]{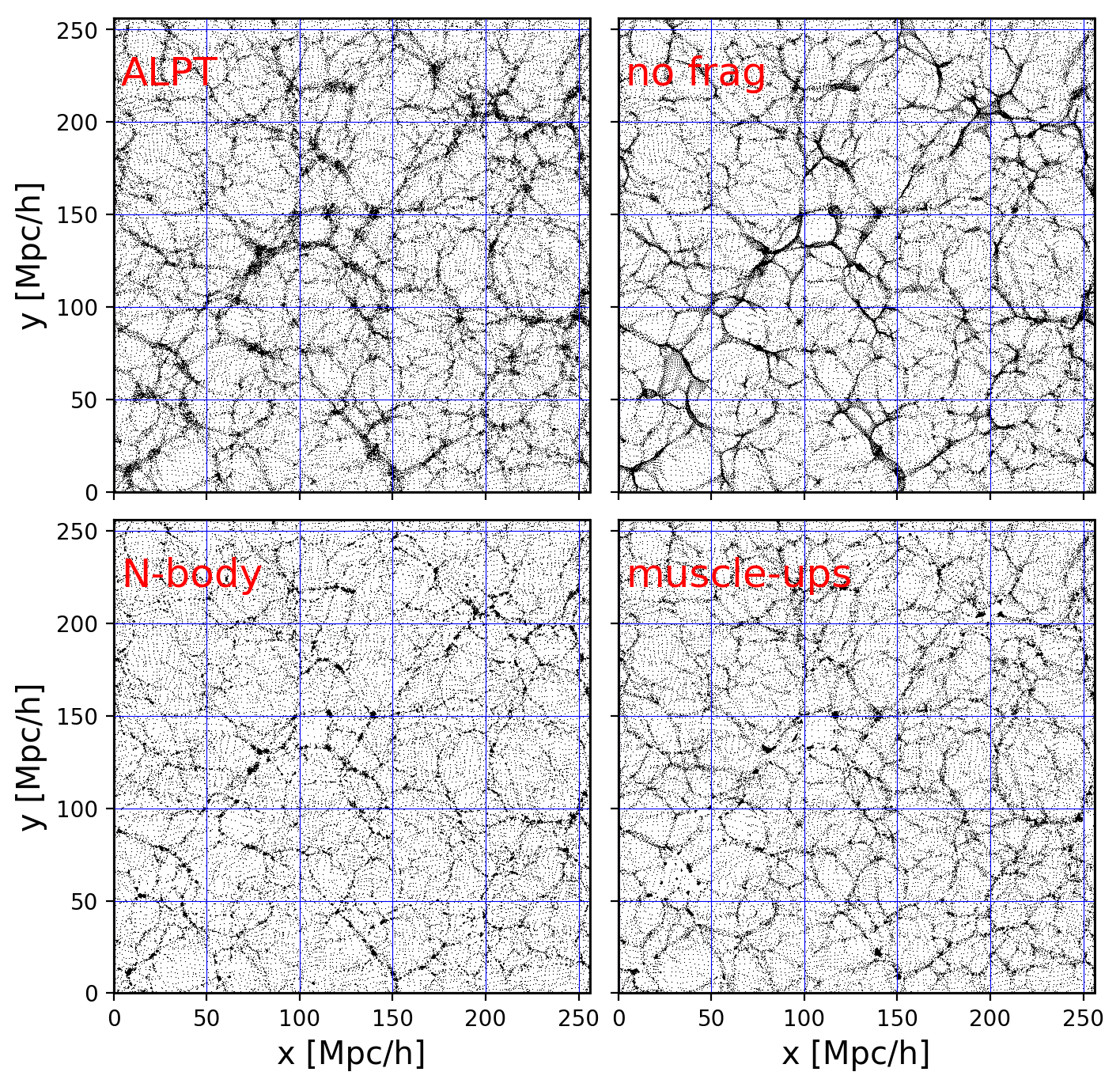}
    \caption{Eulerian positions of the particles in a one-particle-thick Lagrangian slice of the simulations. We plot the \muscleups\ displacement without the fragmentation into haloes in the upper right panel, where one can notice the improvement over ALPT in collapsing filaments and haloes, which we attribute to the increased size of halo patches that make the result more similar to what we obtained thorough \origami. Some of the particles, however, are not collapsed at all. This issue can be corrected by incorporating a halo model (bottom right panel), where these particles are redistributed into NFW haloes as found in the initial conditions.}
    \label{fig:Fig7}
\end{figure*}
We start to examine the results by first looking at the particle displacements in \Cref{fig:Fig7}. In the simpler case where no fragmentation of haloes is implemented (top right panel), we can see the improvement over ALPT, which is reflected in a more defined cosmic web, where haloes and filaments are clearly more compact. This is much more similar to the results obtained through the \textit{a posteriori} simulation based on \origami\ in the bottom left panel of \Cref{fig:Fig2}, where the exact information on halo particles is exploited. The only visible flaw in the non fragmented case is that some regions, especially large dense ones, have particles which are scattered around and collapsed neither to haloes nor filaments. We checked that these particles are in fact halo particles ($\psi=-3$), and this effect might be an artefact of the patch expansion process, which includes a larger fraction of halo particles than $N$-body, which are then not well-fragmented (see \Cref{fig:Fig4}).
\begin{figure*}
    \centering
    \includegraphics[width=.65\textwidth]{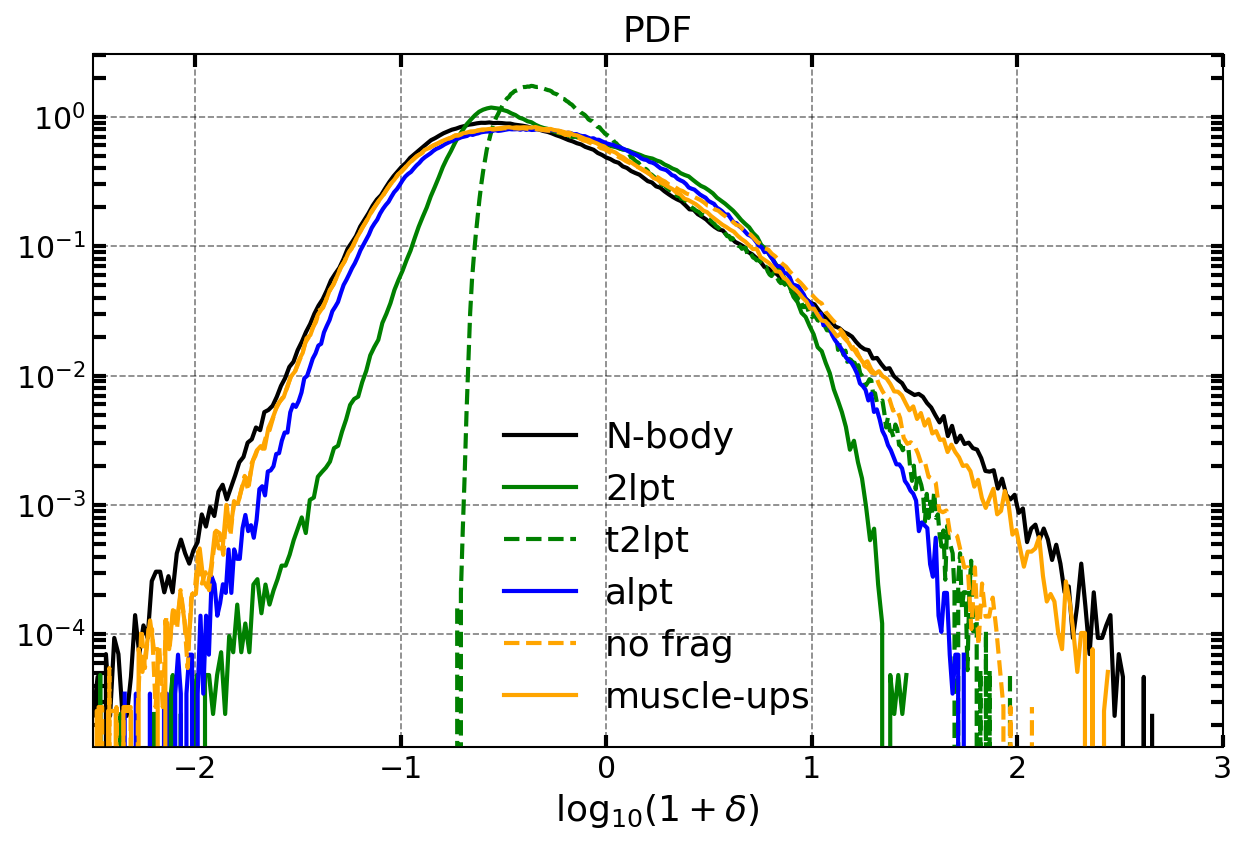}
    \caption{Eulerian PDF of the density field generated through various approximation schemes. The density field has been measured from the $256^3$ particles on a mesh grid of $128^3$ cells, with a cloud-in-cell mass assignment scheme. The density field has been corrected in Fourier space for the effect of the mass assignment with the related cloud-in-cell kernel \protect\citep{Hockney}.}
    \label{fig:Fig8}
\end{figure*}

This is indeed confirmed by the complete \muscleups\ implementation (\Cref{fig:Fig7} bottom right), where this effect disappears, with these extra particles effectively collapsed onto haloes, while the overall cosmic web is not disrupted. It is actually possible to see that there is an excellent correspondence between the knots in the $N$-body and the ones in the halo model approximation.

The effects of the halo fragmentation are also apparent in \Cref{fig:Fig4}, at the level of the displacement field. For the first time, we are able to show that the filamentary structure of the $\psi$ field (not to be confused with density field filaments), is related to the collapse of haloes. In $N$-body, these thin discs between haloes stretch out to form filaments, whereas in \muscleups, they are often only a particle thick, indicating a displacement between haloes that may not be filled in with filament particles between them. While perfect correspondence with the $N$-body displacement field cannot be expected, it is still impressive to see that a toy halo model added \textit{by hand} can qualitatively reproduce this feature. We notice how these $\psi$-filaments are rounder than what is observed in the $N$-body. While we did not impose any sphericity in grouping particles into haloes, it seems that the roundish shape of proto-haloes determined through \Cref{eq:tw3} is mostly preserved through the process.

We also note that, away from the halo core, our NFW algorithm retains the original order of particle distances in \muscleups\ without fragmentation (bottom left panel of \Cref{fig:Fig4}), which produces a smooth $\psi$ field inside each halo (notice that the same occur for the \origami-informed realization). In $N$-body, the scalar displacement field inside a halo patch is rather random, which is made apparent by the noise of $\psi$ around the value $-3$ inside halo-patches.
\begin{figure*}
    \centering
    \includegraphics[width=.95\textwidth]{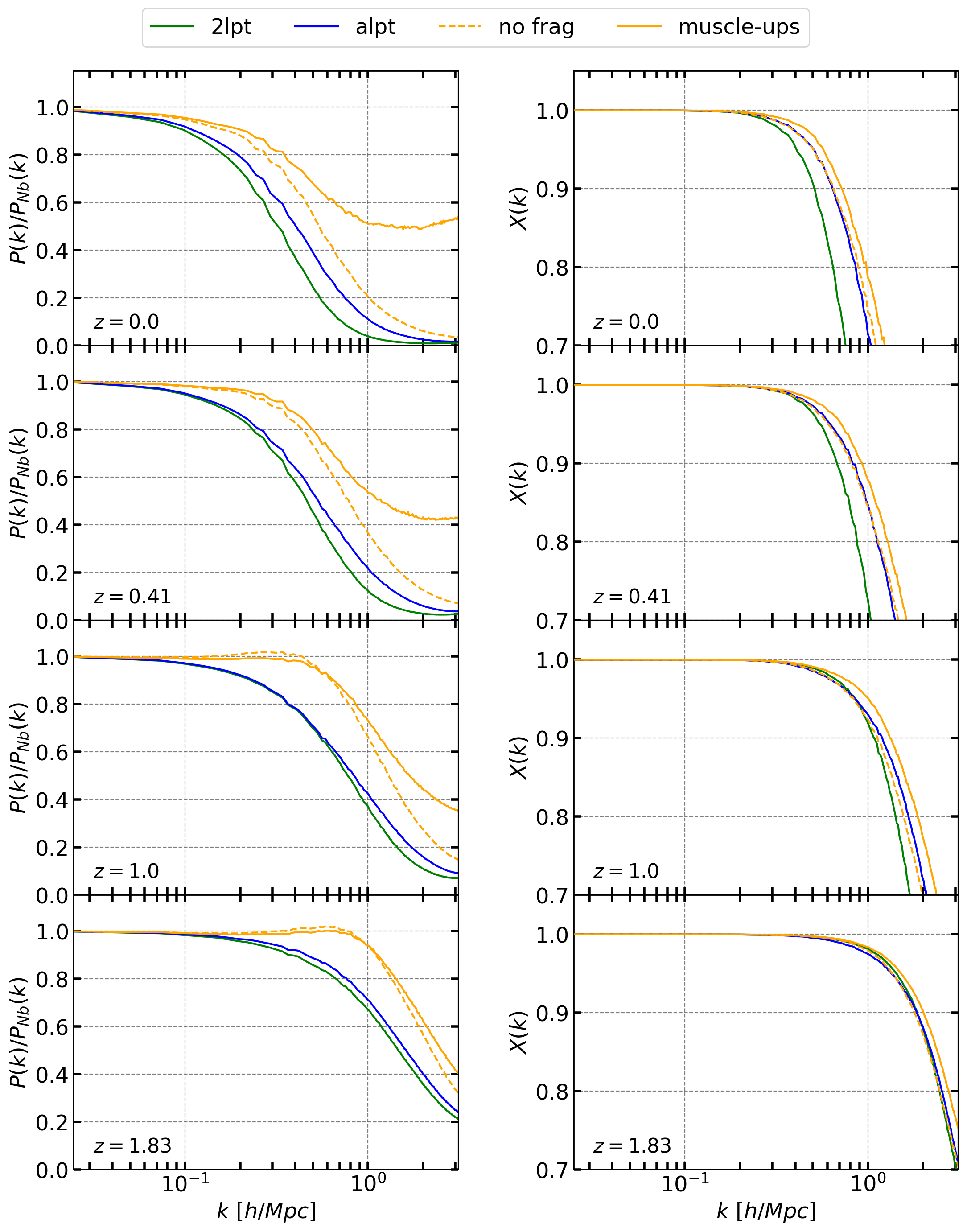}
    \caption{Power spectra and cross-correlations of the density fields generated through ALPT, \muscleups\ and 2LPT, compared to the density field of an $N$-body simulation run with the same initial conditions. The label \textit{no frag} refers to the realization where particles have not been fragmented into haloes, but only displaced according to \Cref{eq:tw1}.
    \label{fig:Fig9}}
\end{figure*}

We now examine the statistics of the density field, starting with the PDF plotted in \Cref{fig:Fig8}. A pure perturbation theory based scheme like 2LPT fails drastically to account for the tails. The PDF is even worse for T2LPT, where the voids are completely unaccounted for. The benefit of SC formula (\Cref{eq:sc}) becomes apparent in ALPT, where the low density tails improves thanks to a better description of voids. \muscleups\ further improves the modelling of the voids, thanks to the multi-scale modelling, but the most apparent gain in using the halo model is visible at the high density tail.

In \Cref{fig:Fig9} we show the results for the cross correlation and the power spectrum statistics at various redshifts. Looking at the cross correlation statistics, we can see that there is a marginal increase at all the redshifts. At $z=0$ and $k=1$ \mpch, we find $0.79$ for our implementation against $0.71$ for ALPT. Although this gain in the cross correlation looks marginal at first glance, it is in fact remarkable; it is much harder to improve this statistics than the power spectrum, as it implies improving the phases too.

Concerning the power spectrum, there is a recovery of power in the quasi linear regime. This recovery is much similar to what we achieved through \origami\ and \rockstar\ \textit{a posteriori} realizations in \Cref{fig:Fig1}. To quote some numbers, at $k=0.1$ ($0.3$) \mpch\ and redshift $z=0$, we find that the accuracy of the matter power spectrum measured from a realization generated through \muscleups is about $5 \%$ ($18 \%$). This is marginally better than ALPT, which has $8 \%$ ($37 \%$). As we go to higher redshifts, the improvement becomes more apparent, most notably at higher $k$; at the same wavenumbers we quote $1 \%$ ($1.1 \%$) against $3 \%$ ($14 \%$) of ALPT.

In both ALPT and \muscleups\ there is an interpolation scale, $\sigma_R$, which is a free parameter. We fix it to $3.0$ \hmpc\ at $z=0.0,0.43$ and $2.5$ and $1.8$ \hmpc\ at $z=1.0, 1.83$ respectively for \muscleups. For ALPT we fix $\sigma_R$ to $3.0$ \hmpc\ at $z=0.0,0.43$ and $2.0$ \hmpc\ at $z=1.0, 1.83$. The value of $\sigma_R$ decreases as we move to higher redshift, since the SC regime starts at the very smallest scales and becomes less relevant. We set this value after trying a handful of them, choosing the one that gives the highest the cross correlation.

As expected, we find a strong dependence of the one-halo term of the power spectrum on the halo-model ingredients. For example, the results we discussed so far have been obtained by using \textit{twice the concentration} of \Cref{eq:cM}. This results in a decrease in the power spectrum for the one-halo term. By using the exact relation instead, we find that haloes are smaller than what is seen in the $N$-body (\Cref{fig:Fig7}), and the resulting power spectrum shows a sharp increase at the transition from large scales to the nonlinear regime, alongside with a decrease of the cross correlation (which becomes comparable to the no-fragmentation implementation of \muscleups). 

We tried to fine-tune the concentration parameter in order to match more precisely the power spectrum at the smaller scales, but we could not find a perfect match. Ultimately this is not a source of concern, first because also the original halo model cannot recover the power of the one halo term, \cite[see the first figure of][for an example]{Mead15}, which could be a hint for the need of more sophisticated concentration parameters and halo profiles than the ones originally implemented. Second, we think that we should weight more the improvement in the cross correlation statistics, which is one of the most difficult to improve, rather than the small-scale power spectrum. The reason is that even if a perfect match of power at small scales is obtained, this is not representative of the real particle positions in the simulation unless both the cross correlation and the power spectrum at the transition between large and small scales are recovered as well.

Another way to affect both $P(k)$ and $X(k)$ is to change the sampling of the target HMF (\Cref{sect:toyhalo}). We find that a finer mass resolution helps to increase $X(k)$ with a slight decrease in $P(k)$. Last, the interpolation scale of ALPT is also relevant for $X(k)$ and $P(k)$ statistics, with a higher scale resulting in a higher power spectrum and a lower cross correlation. In this study we did not optimize the combination of all these parameters, as the results we obtained seem pretty stable under their changes, or affected mostly at the level of the smaller scales.

\section{Summary \& Conclusions}\label{sect:conclusions}
In this work we presented \muscleups, a semi-analytical Lagrangian simulation scheme to approximate the displacement of CDM particles. \muscleups\ is based on the observation that the SC criterion implemented with a Gaussian window detects the innermost regions of proto-halo patches. Since the collapse of particles into haloes is not a standalone process, one should expand these overdense regions to include neighbouring voxels, exploiting a similar concept behind the numerical implementation of the EPS formalism; this prescription increases the resemblance to the $N$-body result (\Cref{fig:Fig4}).

Expanding the halo voxels has the counter effect of increasing the number of halo particles, which in turn increases the linear power spectrum (\Cref{fig:Fig5}). We adopt the same ansatz of ALPT by interpolating the displacement field on small scales with perturbation theory on large scales. Formally, the Lagrangian displacement of \muscleups, without the halo model, is summarized by \Cref{eq:tw1,eq:tw3}.

There is a degree of arbitrariness in how to expand the overdense regions: we find an approach that works well by considering all the particles within a distance $R$ from a collapsing voxel, where $R$ is the scale of the Gaussian filter applied to the linear density field to detect the collapse. We find that the adoption of a Gaussian filter is crucial, as a top-hat window seems to overestimate the collapsing regions, even without expanding them. In addition, the use of a Gaussian window allows us to find many smaller proto-halos when compared to top-hat based EPS, which then can be merged together to match a HMF based on the freedom in the choice of merging events among the many candidate haloes (\Cref{fig:Fig6}). Finally, we used this halo catalogue to implement a toy halo model.

The generation of the displacement field \Cref{eq:tw1} is rather fast, and comparable to \muscle, as most of the computational time is spent to perform the multi-scale smoothing of the initial conditions. However, in its complete form, \muscleups\ has higher complexity than \muscle, as most of the computational time is spent in matching the target HMF by sifting through the halo candidates. This process can be greatly optimized through parallelization by dividing the grid into subgrids, a task that we leave for a future work. This will also include an analysis of the statistics of the halo field that we generate through the halo catalogue building process.

The improvement of \muscleups\ over previous schemes is apparent from the particle displacement (\Cref{fig:Fig7}); one can see that voids are accurately described, the cosmic web is preserved as in $N$-body, and haloes are collapsed. This result is quantified by the PDF (\Cref{fig:Fig8}), the power spectrum and the cross correlation (\Cref{fig:Fig9}) of the density field.

While an exact modelling of the small scale dynamics still eludes us, our approach improves the resemblance to the exact result at least on a statistical level. In fact, for the first time we are able to reproduce the filamentary structure of the displacement field (\Cref{fig:Fig4}), finding that it emerges as a consequence of collapse into haloes.

\section*{Acknowledgements}
FT thanks the Dipartimento di Fisica ``Aldo Pontremoli'' of the University of Milano, and the Dept.\ of Theoretical Physics (now simply the Dept.\ of Physics) at the University of the Basque Country in Bilbao for hospitality during the development of this work. 
LG, BRG and FT acknowledge financial support by grant MIUR PRIN 2017 ``From Darklight to Dark Matter.'' MCN is grateful for funding from Basque Government grant IT956-16. FT is grateful to Enzo Branchini and Carmelita Carbone for useful discussions, and to Pierluigi Monaco for all the useful comments which helped to greatly improve the final manuscript.

\section*{Data availability}
The \muscleups\ code will be shared on reasonable request to the corresponding author.

\bibliographystyle{mnras}
\bibliography{hmuscle.bib}

\appendix
\section{Tidal Field}\label{appendix:tidal}
The connection of tidal fields to the cosmic web has been known for a long time \citep{Bond96}. Tidal effects can be expressed in terms of the eigenvalues of the shear tensor, that we already defined in \Cref{subsect:ZA} from \Cref{eq:masselement}. This tensor can be readily computed in Fourier space 
\begin{equation}
    \partial_{ij}\phi^{(1)} = \mathcal{F}^{-1}\left[ k_i k_j \phi^{(1)}(\pmb{k}) \right],
\end{equation}
where $\mathcal{F}^{-1}$ stands for the inverse Fourier transform. In the ZA limit $\psi^{(1)} \sim -\delta$, so that the displacement potential is
\begin{equation}
    \phi^{(1)}(k) = -\frac{\delta^{(1)}(k)}{k^2},
\end{equation}
and the stress tensor is defined as
\begin{equation}
    \partial_{ij}\phi = \mathcal{F}^{-1}\left[-\frac{k_i k_j}{k^2} \delta^{(1)}(\pmb{k}) \right].
\end{equation}

By diagonalizing the shear tensor matrix at every point $\pmb{q}$, one can use the eigenvalue fields $\lambda_1 \leq \lambda_2 \leq \lambda_3$ instead of the full matrix expression. This is convenient in the linear theory limit, where one can regard the moment where $|\lambda| \sim 1$ as the onset of shell crossing along that particular eigen-direction. Ultimately, this can be used as an approximation to classify the morphology of the cosmic web from the initial conditions \citep{Hahn07}, or to express the anisotropy of the asphericity of gravitational collapse \citep{Sheth02}.
\begin{figure*}
    \centering
    \includegraphics[width=0.8\textwidth]{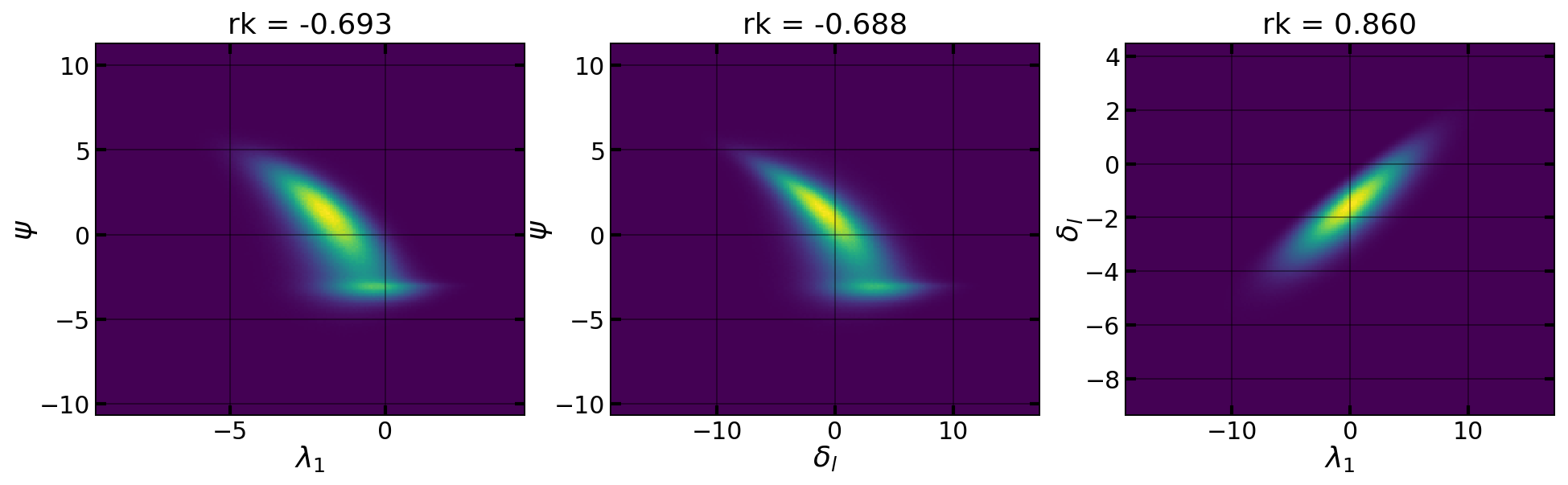}
    \caption{Plots of the correlation between the displacement field and the linear theory eigenvalues. The value on top is the Spearman's rank correlation coefficient.}
    \label{fig:FigA1}
\end{figure*}

We thus analyze the correlations between the eigenvalues and the displacement field in order to see if there is some unexploited information contained in the fields of the eigenvalues. Unfortunately, \Cref{fig:FigA1} shows that there is a high amount of correlation between $\delta_\ell$ and $\lambda_i$, which can be expected since $\delta_\ell = \lambda_1+\lambda_2+\lambda_3$. In fact, we find that it is also possible to define the displacement solely in terms of the first eigenvalue, i.e. a mapping $\psi(\lambda_1)$, but it performs similarly to the approximation schemes discussed in \Cref{sect:theory} based on $\psi(\delta_\ell)$.

\begin{figure*}
    \centering
    \includegraphics[width=0.9\textwidth]{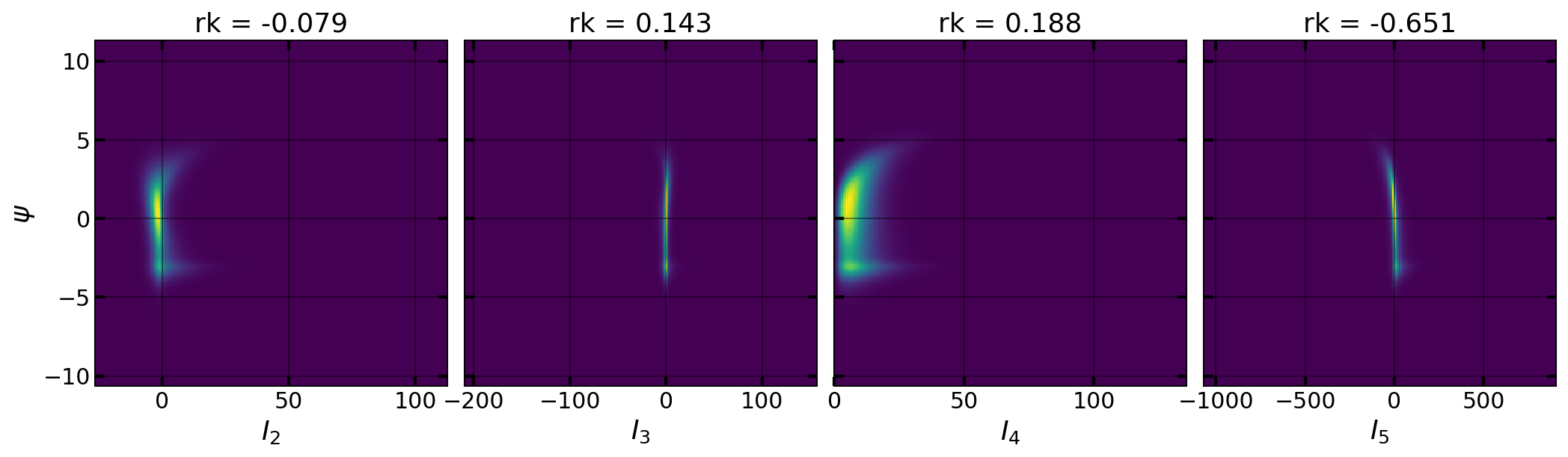}
    \caption{Correlation of the cosmic web invariants with the displacement field $\psi$. The value on top is the Spearman`s rank correlation coefficient. Apart from $I_1=\delta_\ell$, only $I_5$ seems to be correlated with the displacement. Unfortunately the Spearman`s rank correlation between $I_5$ and $\delta_\ell$ is about 0.95, which shows how it does not provide new information with respect to the linear density.}
    \label{fig:FigA2}
\end{figure*}

We do not rule out the possibility that complicated non-linear combinations of the eigenvalues could exist, which may be related to $\psi$ in a non-trivial way. We investigate this scenario by considering the cosmic web invariants (e.g, \cite{Kitaura20})
\begin{equation}
\begin{aligned}
    I_1 &= \lambda_1 + \lambda_2 + \lambda_3,\\
    I_2 &= \lambda_1 \lambda_2 +\lambda_1 \lambda_3+\lambda_2 \lambda_3,\\
    I_3 &= \lambda_1 \lambda_2 \lambda_3,\\
    I_4 &= \lambda^2_1 + \lambda^2_2 + \lambda^2_3,\\
    I_5 &= \lambda^3_1 + \lambda^3_2 + \lambda^3_3.
\end{aligned}
\end{equation}
Cosmic web invariants are fundamental fields which can be used to define all the tidal field related quantities of interest, such as ellipticity, prolateness and fractional tidal anisotropy (e.g. \cite{Paranjape18}). An extensive analysis of cosmic web invariants and their relation to the Eulerian halo field was performed by \cite{Kitaura20}, where they justify them as a way of fully characterizing the cosmic web. We instead consider them in relation to the Lagrangian displacement field, i.e. defined through the linear eigenvalues. From \Cref{fig:FigA2}, one can see how the only significant correlation, apart from the trivial one with $I_1$, is the one with $I_5$. Unfortunately $I_5$ is also highly correlated with $\delta_\ell$ As we already mentioned, defining a map of $\psi$ in terms of any of these quantities does not provide a better modelling than the usual mapping $\psi(\delta_\ell)$, whose limits in the amount of information carried has been thoroughly discussed in \cite{Tosone20}.

These results do not show that the tidal field is not important to consider for modelling the LSS, but they show that the modelling of the scalar displacement field is likely not to gain a major benefit in trying to incorporate these effects, for which reason we omitted them in the main text, and we focused on halo particles alone as a possible way to improve the modelling of displacement field.

\section{Halo Model Ingredients}\label{appendix:halo_ingredients}
For a review about the halo model and its ingredients we refer the reader to \cite{Cooray02}; here we summarize the main quantities we need in the following. The NFW density profile is \citep{Navarro96}
\begin{equation}\label{eq:NFW}
\rho_{\text{nfw}}(r) = \frac{\overline{\rho}_{m}(z) \Delta_v(z) f(c) }{c \frac{r}{r_v}\left( 1+c \frac{r}{r_v} \right)^2},
\end{equation}
where $c$ is the concentration parameter, which can be regarded as a rescaling that modulates the density of the profile and $r_v$ is the virial radius, namely the radius which contains the virial mass $M$ of the halo, corresponding to an overdensity $\Delta_v(z)$ (e.g. $\sim 178$ in Einstein-de Sitter). These quantities must be related through the mass conservation relation
\begin{equation}\label{eq:Mass}
M = \frac{4 \pi}{3} r^3_v(z) \overline{\rho}_{m} \Delta_v(z).
\end{equation}
For the nonlinear spherical overdensity at virialization $\Delta_{v}$, we use the fitting formula by \cite{Bryan&Norman98}
\begin{equation}\label{eq:B&N}
    \Omega_m(z) \Delta_v(z) = 18 \pi^2+82(\Omega_m(z)-1)-39(\Omega_m(z)-1)^2.
\end{equation}

The normalization function $f(c)$ in \Cref{eq:NFW} is derived under the assumption that all the mass of the halo is within the virial radius, thus by integrating \Cref{eq:NFW} and truncating the integration at the virial radius, one can solve for the normalization
\begin{equation}\label{eq:fc}
    f(c) = \frac{c^3}{3 \left( \ln(1+c)-\frac{c}{1+c} \right)}.
\end{equation}

The concentration parameter encompasses important information about the cosmological model and structure formation \citep{Mead16}. In this work we use the median relation found by \cite{Bullock01}
\begin{equation}\label{eq:cM}
c(M) = \frac{9}{1+z}\left(\frac{M}{M_*(z)}\right)^{-0.13},
\end{equation}
where $M_*$ is the critical mass such that $\sigma(M_*,z)=\delta_c$. This formula has been validated for a vanilla $\Lambda CDM$ cosmology, for haloes in the mass range $\sim 10^{11}-10^{14} \ M_{\odot}/h$, thus it adapts well to the present study and to the mass resolution we considered. Despite the halo model in this form is rather raw, and it is already known that there are refined fits for the concentration parameters or halo profiles \citep{Jie19}, we stick to it because of its simplicity. Also, we do not expect that our implementation of the halo model, regardless of the precision of the fitting formulae we adopt, can be precise enough to recover exactly the $N$-body result. It is already known that the halo model is just a simplification of the actual matter distribution, and it cannot recover exactly the same power spectrum as in N-body, unless some phenomenological modifications are applied \citep{Mead15}.

\section{Redistributing Halo Particles}\label{appendix:halo_displacement}
As we said in \Cref{sect:muscleups}, particles are first displaced according to a Lagrangian scheme of our choice, which in our case is set by \Cref{eq:tw1}. Since we already have the halo catalogue, we just need to redistribute particles with the same halo-id  directly in Eulerian space, according to a NFW profile.

For each halo, we first find its barycenter and then use the expected nonlinear overdensity \Cref{eq:B&N} to compute the virial radius from \Cref{eq:Mass}. The halo density profile is then completely specified by \Cref{eq:cM,eq:fc}. To redistribute the particles of each halo around its barycenter according to a NFW profile, we first rank-order the particles based on their distances from the barycenter, which corresponds to the cumulative distribution function (CDF) of the halo profile. Afterwards, the inverse CDF of the NFW density profile is used in order to displace particles according to the desired distribution.

While this can be done in a fully numerical approach, \cite{Robotham18} found out that the inverse CDF of a NFW profile has an analytical form. To see this we define the dimensionless variable $q=r/r_v$ that indicates the distance from the halo center. The mass enclosed within a distance $q$ from the center is
\begin{equation}
    M(q) \propto \ln(1+qc)-\frac{qc}{1+qc}.
\end{equation}

It is convenient to define the cumulative distribution function of the enclosed mass, which is simply $p=M(q)/M(1)$, where the total mass is $M=M(1)$, and to recast the previous equation by adding $1$ on both sides
\begin{equation}\label{eq:preinversion}
    pM(1) +1 = \ln(1+qc)+\frac{1}{1+qc}.
\end{equation}

We exponentiate this equation, and exploit the fact that it is possible to invert it by the means of the Lambert function $W$. In other words the transcendental relation $y=e^ {1/x}$ can be inverted as $x=-1/W(-y^{-1})$. An implementation of the Lambert function is found in the GNU Scientific Library\footnote{\url{https://www.gnu.org/software/gsl/doc/html/specfunc.html}}. By applying it to \Cref{eq:preinversion}, we get \citep{Robotham18}
\begin{equation}
    q = -\frac{1}{c}\left( 1+ \frac{1}{W(-e^{-pM(1)-1})} \right),
\end{equation}
which defines a procedure that allows us to preserve the rank ordering of the halo particles, by mapping each rank $p$ to a distance $q$ from the center.

The previous procedure could reduce the asphericity of the haloes, as they are redistributed as an NFW profile. We then try to preserve the triaxiality by adopting the same scheme used in \cite{Mead14}. The first thing is to compute the inertia tensor of each halo, before its particles are redistributed
\begin{equation}
    I_{ij} = m_p \sum^{N}_{k=1} \left(|\Delta \pmb{x}^{(k)}|^2\delta_{ij}-\Delta x^{(k)}_{i} \Delta x^{(k)}_{j} \right),
\end{equation}
where $\Delta x^{(k)}_{i}$ is the $i$-th component of the $k$-th particle, and $\Delta x^{(k)}_{i} = x_i^{(k)}- \overline{x}_i$ is the distance from the halo barycenter along the $i$-th direction. For each halo, we diagonalize the tensor and save its eigenvalues and eigenvectors. After particles have been distributed to match the NFW profile, we restore the triaxiality by rescaling each $i$-th particle in the halo along its eigendirections by a factor proportional to the eigenvalues
\begin{equation}
\begin{aligned}
    \Delta x_i &= 3a \Delta x_i/(a+b+c),\\
    \Delta y_i &= 3b \Delta y_i/(a+b+c),\\
    \Delta z_i &= 3c \Delta z_i/(a+b+c),
\end{aligned}
\end{equation}
with $a,b,c$ being the square root of the eigenvalues of $I_{ij}$ \citep{Mead14}.

\bsp	
\label{lastpage}
\end{document}